\documentclass[a4paper]{jpconf}
\usepackage{graphicx}
\usepackage{amsmath,amssymb,bm}
\usepackage{color}

\newcommand{\aap}{Astronomy \& Astrophysical}
\newcommand{\apj}{The Astrophysical Journal}
\newcommand{\apjl}{The Astrophysical Journal Letters}
\newcommand{\araa}{Annual Review of Astronomy and Astrophysics}
\newcommand{\grl}{Geophysical research Letters}
\newcommand{\jcap}{Journal of Cosmology and Astroparticle Physics}
\newcommand{\jgr}{Journal of Geophysical Research}

\newcommand{\nat}{Nature}
\newcommand{\planss}{Planetary and Space Science}
\newcommand{\prd}{Physical Review D}
\newcommand{\prl}{Physical Review Letters}

\newcommand{\ssr}{Space Science Review}

\begin{document}
\title{Analyses of residual accelerations for TianQin based on the global MHD simulation}

\author{Wei Su$^{1,2,3}$, Yan Wang$^1$, Ze-Bing Zhou$^1$, Yan-Zheng Bai$^1$, Yang Guo$^3$, Chen Zhou$^4$, Tom Lee$^3$, Ming Wang$^5$, Ming-Yue Zhou$^1$, Tong Shi$^{6}$, Hang Yin$^{1}$, Bu-Tian Zhang$^{1}$}


\address{$^1$ MOE Key Laboratory of Fundamental Physical Quantities Measurements, Hubei Key Laboratory of Gravitation and Quantum Physics, PGMF and School of Physics, Huazhong University of Science and Technology, Wuhan 430074, China}
\address{$^2$ Key Laboratory of Dark Matter and Space Astronomy, Purple Mountain Observatory, Nanjing 210008, China}
\address{$^3$ MOE Key Laboratory for Modern Astronomy and Astrophysics, Nanjing University, Nanjing 210023, China}
\address{$^4$ Department of Space Physics, School of Electronic Information, Wuhan University, Wuhan 430072, China}
\address{$^5$ Institute of Space Weather, School of Mathematics and Statistics, Nanjing University of Information Science and  Technology, Nanjing 210044, China}
\address{$^6$ Department of Climate and Space Sciences and Engineering, University of Michigan, Ann Arbor, MI 48109, USA}

\ead{ywang12@hust.edu.cn, suw12@hust.edu.cn}

\begin{abstract}

TianQin is a proposed space-based gravitational wave observatory. It is designed to detect 
the gravitational wave signals in the frequency range of 0.1 mHz -- 1 Hz. 
At a geocentric distance of $10^5$ km, the plasma in the earth magnetosphere will contribute as the main source of environmental noises. 
Here, we analyze the acceleration noises that are caused by the magnetic field of space plasma 
for the test mass of TianQin. 
The real solar wind data observed by the Advanced Composition Explorer are taken as the input of the magnetohydrodynamic simulation. The Space Weather Modeling Framework is used to simulate 
the global magnetosphere of the earth, from which  
we obtain the plasma and magnetic field parameters on the detector's orbits 
at $\varphi_{s}$ = $0^{\circ}$, $30^{\circ}$, $60^{\circ}$ and $90^{\circ}$,  
where $\varphi_{s}$ is the acute angle between the line that joins the sun and the earth and 
the projection of the normal of the detector's plane on the ecliptic plane. 
We calculate the time series of the residual accelerations and the corresponding 
amplitude spectral densities on these orbit configurations. 
We find that the residual acceleration produced by the interaction between 
the TM's  magnetic moment induced by the space magnetic field and the spacecraft 
magnetic field ($\bm{a}_{\rm M1}$) is the dominant term, 
which can approach $10^{-15}$ m/s$^2$/Hz$^{1/2}$ at $f \approx$ 0.2 mHz 
for the nominal values of the magnetic susceptibility ($\chi_{\rm m} = 10^{-5}$) and 
the magnetic shielding factor ($\xi_{\rm m} = 10$) of the test mass. 
The ratios between the amplitude spectral density of the acceleration noise 
caused by the space magnetic field and the preliminary goal of 
the inertial sensor are 0.38 and 0.08 at 1 mHz and 10 mHz, respectively. 
We discuss the further reduction of this acceleration noise by decreasing $\chi_{\rm m}$ 
and/or increasing $\xi_{\rm m}$ in the future instrumentation development for TianQin.

\end{abstract}

\vspace{2pc} \noindent{\it Keywords}: gravitational waves, space plasma, test mass, acceleration noise

\section{Introduction}\label{intro}

Since the direct detection of the gravitational waves (GWs) from the merger of a pair of stellar mass 
black holes (GW150914) by the two advanced detectors 
of the Laser Interferometer Gravitational-Wave Observatory (LIGO) \cite{LIGO2016}, 
more than ten GW events have been detected by advanced LIGO and advanced 
Virgo \cite{2016PhRvL.116x1103A, 2017PhRvL.119n1101A, 2017PhRvL.118v1101A, 
2017ApJ...851L..35A, 2017PhRvL.119p1101A, 2019PhRvX...9c1040A}. 
Recently, KAGRA \cite{Somiya_2012} has joined the ground-based GW detector network. 
Due to the unshieldable impacts from the environment, namely seismic noise and gravity gradient noise, 
it is very difficult for these terrestrial laser interferometers to detect GWs with the frequencies lower than 10 Hz. 
However, in the low frequencies, there are rich sources of GWs that can be used to study 
the fundamental physics, astrophysics and cosmology \cite{2009LRR....12....2S}. 
The aim of space-borne laser interferometers is to explore the GWs in the millihertz range (0.1 mHz--1 Hz). 
Several projects, e.g., LISA \cite{LISA2017}, TianQin (TQ) \cite{Luo2016}, 
DECIGO \cite{2011CQGra..28i4011K}, ASTROD-GW \cite{1998grco.conf..309N}, g-LISA \cite{Tinto2015gLISA}, 
Taiji (ALIA descoped) \cite{TaiJi2015} and BBO \cite{Cutler2006BBO} have been proposed 
and are currently under different stages of study and development.

Both LISA and TQ have three drag-free spacecraft which compose a nearly equilateral triangular constellation. 
Different from LISA, TQ's spacecraft will be deployed in a geocentric orbit with an 
altitude of $1\times 10^5$ km from the geocenter, which makes the distances between each pair of 
spacecraft $\approx$ $1.7\times10^5$ km \cite{Luo2016}. 
The normal of the detector's plane formed by the three spacecraft points toward the candidate 
ultracompact white-dwarf binary RX J0806.3+1527 \cite{Israel2002}. 
The three spacecraft are interconnected by infrared laser beams and form up to three 
Michelson-type interferometers. The heterodyne transponder-type laser interferometers are used to measure the 
displacements of the test masses (TMs) to the accuracy of $10^{-12}~\rm{m/Hz^{1/2}}$ in millihertz. 
The disturbance reduction system is designed to reduce the non-conservative acceleration 
of each TM (Pt-Au alloy, $m = 2.45~\rm{kg}$) 
down to $10^{-15}~\rm{m/s^2/Hz^{1/2}}$ in millihertz \cite{Luo2016}. 
The nominal orbit and a set of alternatives for TQ have been optimized such that the stability 
of the orbits, in terms of the variations of arm lengths, breathing angles and relative range rates, 
can meet the requirements imposed by the long range space laser interferometry \cite{2019IJMPD..2850121Y}. 
The response of TQ as a Michelson interferometer and its sensitivity curve have been given \cite{Hu2018}. 
Its science objectives involving astrophysical sources \cite{2019PhRvD..99l3002F, 2019PhRvD.100d3003W, 
2019PhRvD.100d4036S, 2019PhRvD.100h4024B} and cosmological sources 
\cite{2018JCAP...07..007D,2019arXiv190701838W} are currently under intensive investigations.

In space, the space plasma will serve as the main source of environmental noises. 
One example is the dispersion effect induced by the space plasma 
when the laser beams propagate from one spacecraft to the other \cite{Hutchinson2005}. 
It can cause time-varying optical paths and time delays, hence affects the displacement 
measurement accuracy. 
Taking the typical magnitudes of electron number density and magnetic field in the earth magnetosphere 
as 1--10 cm$^{-3}$ and 1--10 nT \cite{Huang2018,Zhang2007}, respectively, at the altitude of TQ's spacecraft, 
one can see that the dominating factor of the dispersion is from the former.

On the other hand, 
the magnetic field plays a central role in the formation of the structures 
in heliophysics, such as the photosphere and corona of the Sun \cite{Guo2012,Su2018}, 
Parker spiral field lines \cite{Parker1958}, the boundary of the heliosphere \cite{Decker2005}, 
and the magnetosphere of the Earth, Jupiter and Saturn \cite{Bagenal2011,Wang2016}. 
For LISA, it has been shown that the magnetic field of space plasma is the main source of the non-conservative forces 
acting on the TMs housed in the spacecraft through the interactions with its magnetic remanence and 
residual charges \cite{Schumaker2003,Stebbins2004}. 
Evaluating the impacts of these residual accelerations for TQ is the focus of this work.

In Fig.~\ref{fig:0}, TQ's orbit (red circle) will pass through different background environments, including solar wind, bow shock, 
magnetosheath, magnetopause, lobes of magnetosphere, magnetotail and so on. 
There are abundant phenomena of space plasma at this orbit altitude, 
which cover three spatial scales: the global scale, the magnetohydrodynamic (MHD) scale and the plasma scale.
In the global scale, the earth magnetosphere is formed by the interaction between the solar wind and the magnetosphere. The variations of the parameters (e.g., velocity, magnetic field) in solar wind can change the shape and position of the bow shock and the magnetopause \cite{Lv2015}. In the MHD scale, there are instabilities, such as Kelvin-Helmholtz instability at the magnetopause \cite{Hasegawa2004}. In the plasma scale, the turbulance in solar wind and magnetosheath \cite{He2012,Sahraoui2013,Huang2018} and the plasma waves in solar wind and magnetosphere \cite{Li2009,Schwartz2011,Allen2015} are ubiquitous. Besides, the disturbances from the sun (e.g., coronal mass ejections, coronal shocks \cite{Cheng2018,Su2015,Su2016}) and the magnetosphere (e.g., magnetic storms, magnetic reconnections \cite{Huang2012,Takahashi2018,Zhou2019}) take place occasionally. 
All these phenomena can lead to the variations of the electron number density and the magnetic field, 
which in turn lead to the fluctuations of the dispersion along the optical paths and the non-conservative forces 
on the TMs.

The rest of the paper is organized as follows. 
Section \ref{s:2} gives the formalism that is subsequently used to analyze 
the acceleration noises of TMs induced by the magnetic field of the space plasma. 
In section \ref{s:3}, we use a well established MHD model (Space Weather Modeling Framework; SWMF \cite{Toth2005}) to simulate a magnetospheric environment that TQ's spacecraft may encounter. 
The inclinations of TQ detector's plane  
in the magnetosphere ($\varphi_{s}$) are discussed. 
In section \ref{s:4}, based on the simulated magnetic field, we calculate 
the acceleration noises and the associated amplitude spectral densities (ASDs)
at four typical values of $\varphi_{s}$, 
and discuss its reduction in the parameter space of $\xi_{\rm m}-\chi_{\rm m}$. 
The paper is concluded in section \ref{s:5}.

\section{Acceleration noise analysis}
\label{s:2}

In the magnetic field of space plasma, there are two kinds of non-conservative forces acting on the TMs. The first one is the magnetic force produced by the interaction between the TM with a magnetic moment and the background magnetic field. The second one is the Lorentz force that produced by the movement of the TM with residual charges in this magnetic field.  

\subsection{Magnetic force}

The magnetic force on the TM with a magnetic moment ($\bm{M}_{\rm tm}$) in the background magnetic field ($\bm{B}$) can be written as:
\begin{equation}
\label{equation1}
\bm{F} = \nabla (\bm{M}_{\rm tm} \cdot \bm{B})  \,.
\end{equation}
Here, $\bm{B}$ is composed of the space magnetic field $\bm{B}_{\rm sp}$ and 
the spacecraft magnetic field $\bm{B}_{\rm sc}$ inherited from the payloads 
(e.g., permanent magnet used in attitude control or laser frequency stabilization), i.e.,  
$\bm{B} = \bm{B}_{\rm sp} + \bm{B}_{\rm sc}$. 
$\bm{M}_{\rm tm}$ is composed of the remanent magnetic moment $\bm{M}_{\rm r}$ 
and the inductive magnetic moment $\bm{M}_{\rm i}$: $\bm{M}_{\rm tm} = \bm{M}_{\rm r} + \bm{M}_{\rm i}$. 
$\bm{M}_{\rm i}$ can be induced both by $\bm{B}_{\rm sp}$ and $\bm{B}_{\rm sc}$,  
\begin{equation}
\label{equation2}
\bm{M}_{\rm i} = \frac{\chi_{\rm m} V_{\rm tm} (\bm{B}_{\rm sp} + \bm{B}_{\rm sc} ) }{\mu_0} = \bm{M}_{\rm isp} + \bm{M}_{\rm isc} \,,
\end{equation}
where, $\mu_0$ is the vacuum magnetic permeability, $\chi_{\rm m}$ is the magnetic susceptibility, 
and $V_{\rm tm}$ is the volume of the TM. 
Inserting $\bm{M}_{\rm tm}$ and Equation (\ref{equation2}) into Equation (\ref{equation1}), the acceleration can be expressed as follows: 
\begin{equation}
\label{equation3}
\begin{split}
\bm{a}_{\rm M} &= \frac{1}{m} \nabla [(\bm{M}_{\rm r} + \bm{M}_{\rm isp} + \bm{M}_{\rm isc} ) \cdot (\bm{B}_{\rm sp}  + \bm{B}_{\rm sc} )] \\
&= \frac{1}{m} \nabla (\bm{M}_{\rm r} \cdot \bm{B}_{\rm sp}  + \bm{M}_{\rm r}\cdot \bm{B}_{\rm sc} 
+ \frac{\chi_{\rm m} V_{\rm tm}}{\mu_0} B_{\rm sp}^2 + \frac{2 \chi_{\rm m} V_{\rm tm}}{\mu_0} \bm{B}_{\rm sp} \cdot \bm{B}_{\rm sc} + \frac{\chi_{\rm m} V_{\rm tm}}{\mu_0} B_{\rm sc}^2) \,,
\end{split}
\end{equation}
where $m$ is the mass of the TM, $B_{\rm sc} = |\bm{B}_{\rm sc}| $, $B_{\rm sp} = |\bm{B}_{\rm sp}| $. 
Based on the vector operation rules, the first term in the second line of Equation (\ref{equation3}) can be expanded as:
\begin{equation}
\label{equation4}
\nabla (\bm{M}_{\rm r} \cdot \bm{B}_{\rm sp} ) = (\bm{M}_{\rm r} \cdot \nabla) \bm{B}_{\rm sp}  + (\bm{B}_{\rm sp} \cdot \nabla) \bm{M}_{\rm r}
+ \bm{M}_{\rm r} \times (\nabla \times \bm{B}_{\rm sp}  ) + \bm{B}_{\rm sp}  \times (\nabla \times \bm{M}_{\rm r} )  \,.
\end{equation}
According to Maxwell Equations, $\nabla \times \bm{B}$ can be induced by the electric current density $\bm{j}$ and the displacement current density $\varepsilon_0 \partial \bm{E} / \partial t$, where $\varepsilon_0 $ is the vacuum electric permittivity. Since the TM is encapsulated by the house of the inertial sensor in the disturbance reduction system, the electric current on the TM can be ignored. Therefore, Equation (\ref{equation4}) can be written as:
\begin{equation}
\label{equation5}
\nabla (\bm{M}_{\rm r} \cdot \bm{B}_{\rm sp} ) = (\bm{M}_{\rm r} \cdot \nabla) \bm{B}_{\rm sp} + (\bm{B}_{\rm sp} \cdot \nabla) \bm{M}_{\rm r}
+ \bm{M}_{\rm r} \times (\frac{\varepsilon_0 \mu_0 \partial \bm{E}_{\rm sp} } {\partial t}) + \bm{B}_{\rm sp} \times (\nabla \times \bm{M}_{\rm r} ) \,.
\end{equation}
Similarly, the second term in Equation (\ref{equation3}) can be expanded as Equation (\ref{equation5}) 
with $\bm{B}_{\rm sp}$ and $\bm{E}_{\rm sp}$  replaced by $\bm{B}_{\rm sc}$ and $\bm{E}_{\rm sc}$. 
And the forth term in Equation (\ref{equation3}) can be expanded as:
\begin{equation}
\label{equation7}
\nabla (\bm{B}_{\rm sp} \cdot \bm{B}_{\rm sc} ) 
= (\bm{B}_{\rm sp} \cdot \nabla) \bm{B}_{\rm sc} + (\bm{B}_{\rm sc} \cdot \nabla) \bm{B}_{\rm sp}
+ \bm{B}_{\rm sp} \times (\frac{\varepsilon_0 \mu_0 \partial \bm{E}_{\rm sc} }{\partial t} ) 
+ \bm{B}_{\rm sc} \times (\frac{\varepsilon_0 \mu_0 \partial \bm{E}_{\rm sp} }{\partial t} ) \,.
\end{equation}
According to $\nabla (uv) = u \nabla v + v \nabla u$, where $v$ and $u$ are scalars, the third and fifth term in Equation (\ref{equation3}) can be written as:
\begin{equation}
\label{equation8}
\nabla (B_{\rm sp}^2) = 2 B_{\rm sp} \nabla B_{\rm sp} \,,
\end{equation}
\begin{equation}
\label{equation9}
\nabla (B_{\rm sc}^2) = 2 B_{\rm sc} \nabla B_{\rm sc} \,.
\end{equation}

Insert Equations (\ref{equation5})--(\ref{equation9}) into Equation (\ref{equation3}), we get: 
\begin{equation}
\label{equation10}
\begin{aligned}
\bm{a}_{\rm M} &= \frac{1}{m} [(\bm{M}_{\rm r} \cdot \nabla) + (2 \bm{M}_{\rm isp} \cdot \nabla) ] \bm{B}_{\rm sc}
+ \frac{1}{m} [(\bm{M}_{\rm r} \cdot \nabla) + (2 \bm{M}_{\rm isc} \cdot \nabla) ] \bm{B_{\rm sp}} \\
&\quad + \frac{1}{m} [(\bm{M}_{\rm r} + 2\bm{M}_{\rm isp} ) \times (\frac{\varepsilon_0 \mu_0 \partial \bm{E}_{\rm sc}}{\partial t} )]
+ \frac{1}{m} [(\bm{M}_{\rm r} + 2\bm{M}_{\rm isc}) \times (\frac{\varepsilon_0 \mu_0 \partial \bm{E_{\rm sp}}}{\partial t} )]\\
&\quad + [(\bm{B}_{\rm sp} + \bm{B}_{\rm sc} ) \cdot \nabla] \bm{M}_{\rm r} + (\bm{B}_{\rm sp} + \bm{B}_{\rm sc} ) \times (\nabla \times \bm{M}_{\rm r} ) \\
&\quad + \frac{1}{m} 2 M_{\rm isc} \nabla B_{\rm sc} + \frac{1}{m} 2 M_{\rm isp} \nabla B_{\rm sp} \,.
\end{aligned}
\end{equation}
The remanent magnetic moment $\bm{M}_{\rm r}$ is approximated as uniform here. 
Since the house of the TM can provide magnetic and electric shielding, so that each magnetic 
and electric field in Equation (\ref{equation10}) needs to be divided by a shielding factor $\xi_{\rm m}$ ($\xi_e$). 
Thus, Equation (\ref{equation10}) can be reorganized as the following six terms: 
\begin{equation}
\label{equation11}
\left\{
\begin{aligned}
\bm{a}_{\rm M1} &= \frac{1}{m \xi_{\rm m}} [(\bm{M}_{\rm r} + 2 \bm{M}_{\rm isp}) \cdot \nabla] \bm{B}_{\rm sc}   \,, \\
\bm{a}_{\rm M2} &= \frac{1}{m \xi_{\rm m}} [(\bm{M}_{\rm r} + 2 \bm{M}_{\rm isc}) \cdot \nabla] \bm{B}_{\rm sp}   \,, \\
\bm{a}_{\rm M3} &= \frac{1}{m \xi_e}(\bm{M}_{\rm r} + 2 \bm{M}_{\rm isc}) \times \frac{\varepsilon_0 \mu_0 \partial \bm{E}_{\rm sp}}{\partial t}  \,, \\
\bm{a}_{\rm M4} &= \frac{2}{m \xi_{\rm m}} M_{\rm isp} \nabla B_{\rm sp}   \,, \\
\bm{a}_{\rm M5} &= \frac{1}{m \xi_e}(\bm{M}_{\rm r} + 2 \bm{M}_{\rm isp}) \times \frac{\varepsilon_0 \mu_0 \partial \bm{E}_{\rm sc}}{\partial t} \,,  \\
\bm{a}_{\rm M6} &= \frac{2}{m \xi_{\rm m}} M_{\rm isc} \nabla B_{\rm sc}  \,. 
\end{aligned}
\right.
\end{equation}

\subsection{Lorentz force}

Energetic charged particles, such as solar energetic particles (SEPs) and galactic cosmic rays (GCRs), can penetrate the shields and make the TMs charged \cite{Wass2005}. The interaction between the resulting charges and the background magnetic field can lead to Lorentz force. 
On the other hand, the electric force induced by the Hall voltage can partially compensate for the Lorentz force on the TM as its metallic enclosure passes through the space magnetic field and effectively serves as a shield.
Following \cite{Sumner2020}, we introduce an effective shielding coefficient $\eta$ in our calculation. 
Thus, 
the acceleration caused by Lorentz force on the TM with residual charge ($q$) in space magnetic field ($\bm{B}_{\rm sp}$) is as follows: 
\begin{equation}
\label{equation13}
\bm{a}_{\rm L} = \frac{\eta}{m} q \bm{v} \times \bm{B}_{\rm sp} \,,
\end{equation}
where $\bm{v}$ is the velocity of the TM and $\eta$ = 0.03 \cite{Sumner2020}.

\subsection{Evaluating residual accelerations}
\label{s:2.3}

As shown in Equation (\ref{equation11}), $\bm{a}_{\rm M1}$ and $\bm{a}_{\rm M2}$ can be simplified 
as the form $(\mathbf{A} \cdot \nabla) \bm{B}$, here $\bm{A}$ and $\bm{B}$ are vectors. 
Since the spatial resolution of the MHD simulation used here is 0.25 $\rm R_E$ (the radius of the Earth) 
which is much larger than the length of the TM (5 cm) \cite{Luo2016}, 
the spatial variation of the direction of $\bm{B}_{\rm sp}$ at the length scale of the TM can be neglected. 
For $(\mathbf{A} \cdot \nabla) \bm{B}$, we have 
\begin{equation}
\label{equation14}
\left| (\mathbf{A} \cdot \nabla) \bm{B} \right| = \left| (A_x\frac{\partial \bm{B}}{\partial x} + A_y\frac{\partial \bm{B}}{\partial y} + A_z \frac{\partial \bm{B}}{\partial z} ) \right|
= \left| (\mathbf{A} \cdot \nabla B)~\bm{e}_{\rm B} \right|
= \left| A \right| \left| \nabla B \right| \cos \theta 
\leq \left| A \right| \left| \nabla {B} \right|,
\end{equation}
where $\bm{e}_{\rm B}$ is the unit vector of $\bm{B}$. Equation (\ref{equation14}) is used here to 
estimate the maxima of $\bm{a}_{\rm M1}$ and $\bm{a}_{\rm M2}$  
and we ignored the spatial variation of the direction of $\bm{B}_{\rm sp}$.

The magnetic field of the permanent magnet onboard in the spacecraft is simplified as a dipole field $\bm{B}_{\rm sc}$ which, at the location of the TM, is 
\begin{equation}
\label{equation15}
\bm{B}_{\rm sc} = -\frac{\mu_0}{4\pi} (\bm{M}_{\rm sc} \cdot \nabla) \frac{\bm{r}}{r^3} \approx 2 \frac{\mu_0 M_{\rm sc} \bm{r} }{4 \pi r^4} \,,
\end{equation}
where the distance between the TM and the permanent magnet $r= 0.5$ m \cite{Luo2016}. 
$\bm{M}_{\rm sc} = 1~\mathrm{A~m^2}$ \cite{Schumaker2003}. The gradient of $B_{\rm sc}$ is 
\begin{equation}
\label{equation16}
\left| \nabla B_{\rm sc} \right| = \left| 2 \nabla ( \frac{\mu_0 M_{\rm sc} \bm{r} }{4 \pi r^4} ) \right| \approx \frac{3B_{\rm sc}}{r} \,.
\end{equation}

There are bulk flow velocity ($\bm{v}$) and magnetic field but no electric field data in the outputs of our simulation. The electric field of magnetosphere ($\bm{E}_{\rm sp}$) can be approximated as $- \bm{v} \times \bm{B}_{\rm sp}$ \cite{Mozer1978}, which is used to calculate $\bm{a}_{\rm M3}$. 

In the following calculations, the mass of the TM is $m = 2.45~\mathrm{kg}$, the side length of the cubic TM is $5 ~\mathrm{cm}$, and $|\bm{M}_{\rm r}| = 20~ \mathrm{nA~m^2}$ \cite{Stebbins2004}. The shielding factors $\xi_{\rm m}$ and $\xi_e$ are both set to be 10 as fiducial values \cite{LPF2011}.  
The magnetic susceptibility is $\chi_{\rm m}= 10^{-5}$ for the Pt-Au alloy TMs of TQ \cite{Luo2016}. 
The residual charge is $q$ = $10^{7} \times 1.6 \times 10^{-19}$ C \cite{LPF2011}.

\section{MHD simulation}
\label{s:3}

\subsection{Global magnetosphere model and input data}
\label{s:3.1}

\begin{figure}[!h]
	\centering
	\includegraphics[width=10 cm, trim = 120 50 120 50]{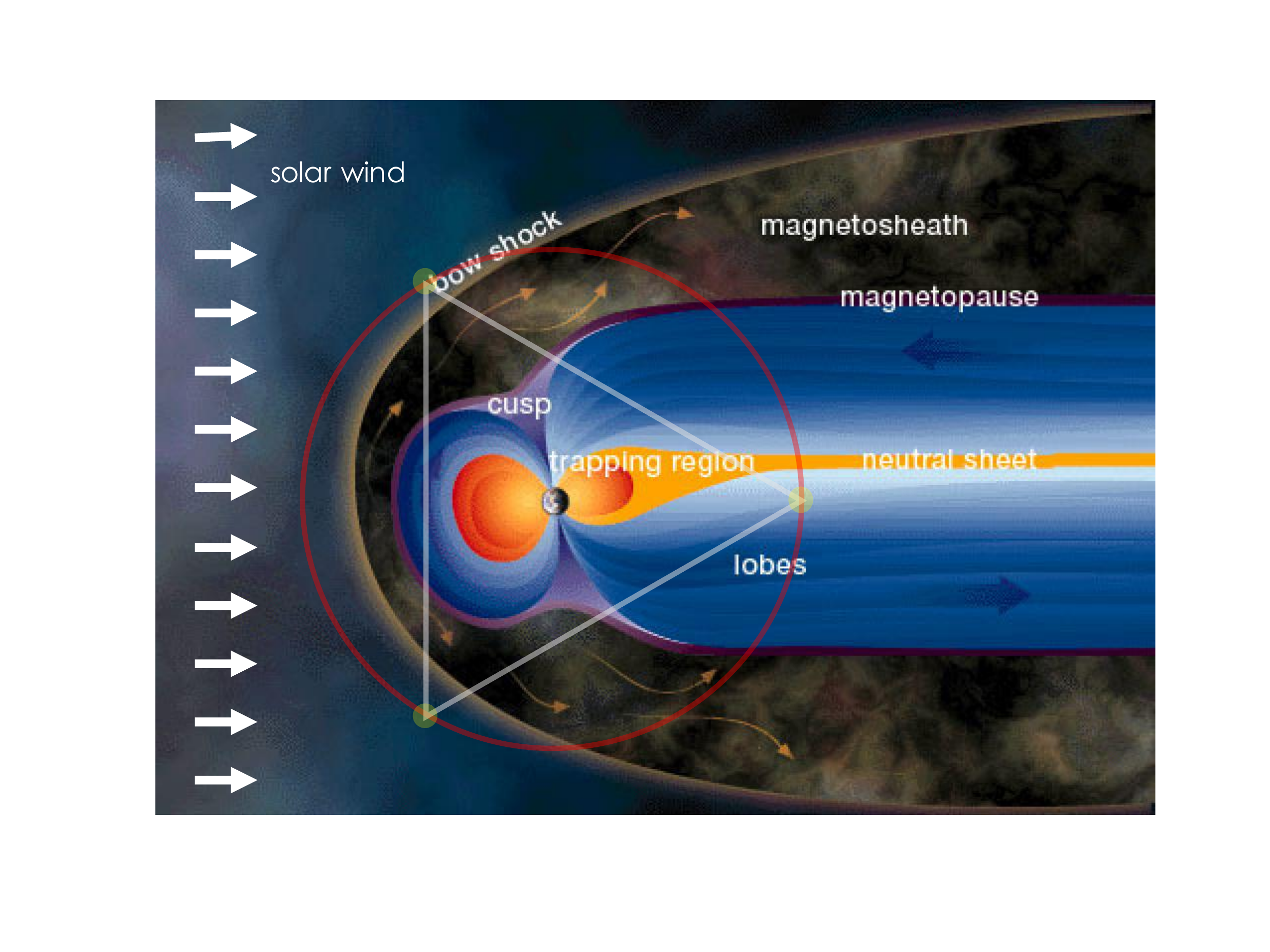} 
	\caption{The schematic view of TQ's orbit (red circle) in the background of the global magnetosphere at $\varphi_{s} \approx 90^{\circ}$. 
	Thin white lines are the laser beams that interconnect the spacecraft of TQ (filled yellow circles). Modified from \cite{Sharma2008}. }
	\label{fig:0}
\end{figure}

We adopt the Space Weather Modeling Framework (SWMF) \cite{Toth2005,Toth2012} to simulate the distributions of 
the parameters (e.g., $\bm{B}_{\rm sp}$) in the space region enclosing the TQ's orbit (Fig.~\ref{fig:0}). 
The SWMF can simulate the interaction between the solar wind and the magnetosphere of the earth. 
It has been used in the studies of magnetospheric physics \cite{Takahashi2018,Lv2015,Wang2016}, and has been thoroughly validated \cite{Welling2010,Dimmock2013,Liemohn2018}.

\begin{figure}[!h]
	\centering
	\includegraphics[width=0.8 \textwidth]{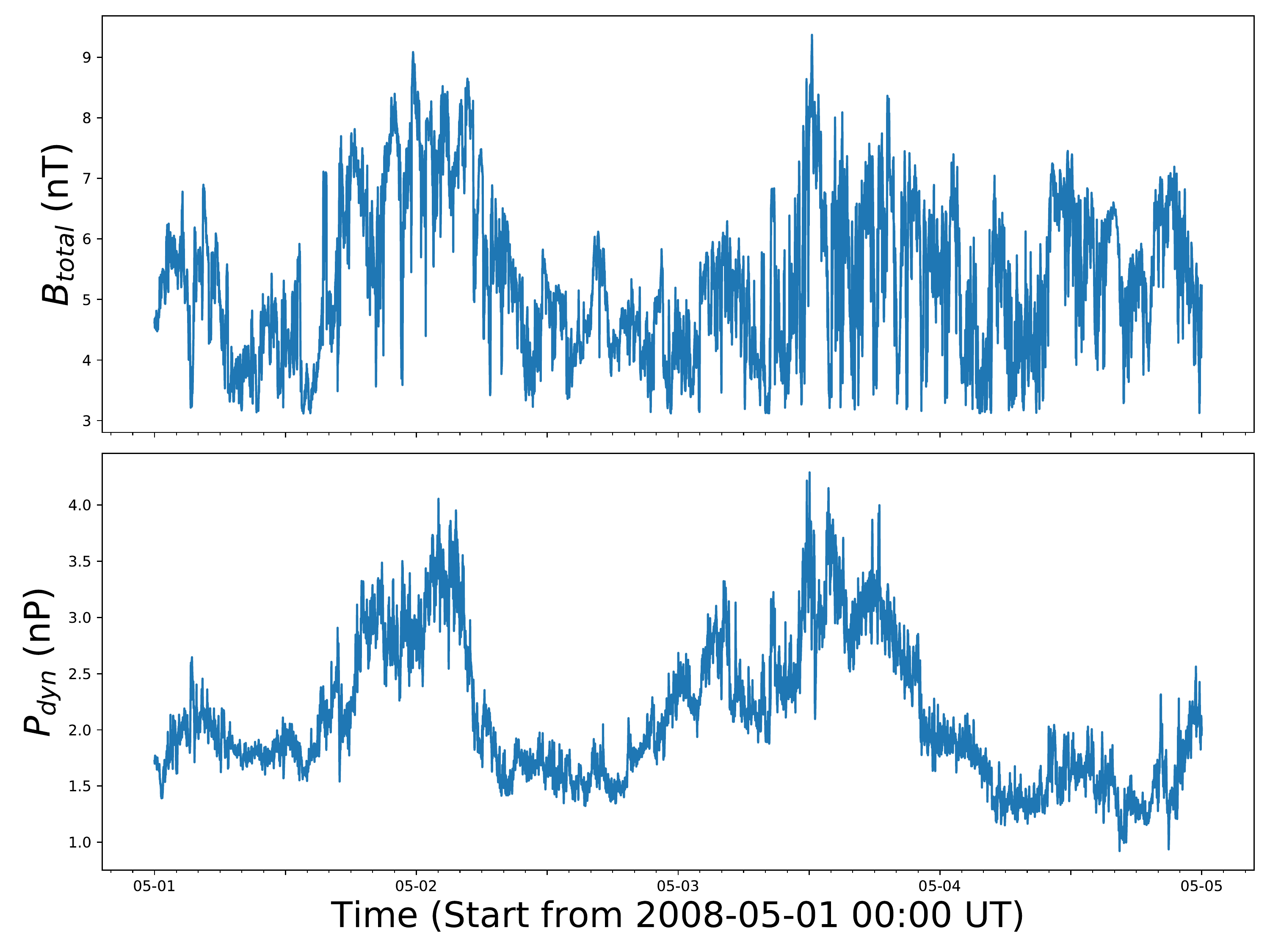}
	\caption{
		The input total space magnetic field ($\bm{B}_{\rm total}$) that is observed by the ACE.
		The solar wind dynamic pressure $P_{\rm dyn}$ is derived from the observation of the ACE. }
	\label{fig:01}
\end{figure}

The SWMF is integrated by several modules including Solar Corona, Inner Heliosphere, Global Magnetosphere, Inner Magnetosphere, etc. 
In this work, we requested our simulation on the Community Coordinated Modeling Center (CCMC) using SWMF/Block-Adaptive-Tree-Solarwind-Roe-Upwind-Scheme (BATSRUS), coupled with the Rice Convection Model (RCM; \cite{Harel1981,Toffoletto2003}) and the Fok Radiation Belt Environment model (RBE; \cite{Fok1997,Fok2008,Glocer2009}) . 
The inputs of the simulation are the real time plasma data that observed by the Advanced Composition Explorer (ACE; \cite{Stone1998})  with a 1-minute cadence, in the time range from 2008-05-01 00:00 UT to 2008-05-04 24:00 UT. 
The input parameters, such as the total space magnetic field $B_{\rm total}$ and the solar wind dynamic pressure 
$P_{\rm dyn}$, are shown in Fig. \ref{fig:01}. 
Here, $P_{\rm dyn}$ is the most important parameter to construct the structure of the earth magnetosphere,  
and it correlates reasonably well with the solar activity cycles \cite{Samsonov2019}. 
From the data of OMNIWeb \cite{King2005}, the mean value of $P_{\rm dyn}$ is 2.0 $\pm$ 1.2 nP 
during 1997 -- 2019 (one total solar cycle). The mean value of $P_{\rm dyn}$ adopted here is 2.1 $\pm$ 0.7 nP, 
which is considered as moderate and typical. 
The simulation domain is defined as $-250~{\rm R_{E}} < x < 33~{\rm R_{E}}$ and $\left| y \right|= \left| z \right| < 48~{\rm R_{E}}$  in the Geocentric Solar Magnetospheric (GSM) coordinates, considering the geocentric distance of each TQ's spacecraft is 10$^5$ km ($\approx16~{\rm R_{E}}$). On the orbit (e.g., the red circle in Fig.~\ref{fig:0}), the finest grids are in the vicinity of the near-tail and the dayside magnetopause with a resolution of 0.25~${\rm R_{E}}$, and the rest has a resolution of 0.5~${\rm R_{E}}$. The outputs contain the plasma parameters (number density of ion $n_i$, electron $n_e$, pressure $P$, bulk flow velocity $v_x$, $v_y$, $v_z$), magnetic field ($B_x$, $B_y$, $B_z$), and electric current ($J_x$, $J_y$, $J_z$) on the grid of the simulation domain. Note that these parameters are in the GSM coordinates which need to be converted to the Geocentric Solar Ecliptic (GSE) coordinates when calculating the acceleration noises in section~\ref{s:4}.

\subsection{Relative positions} 
\label{s:3.2}

\begin{figure}[!h]
	\centering
	\includegraphics[width=0.75\textwidth, trim = 0 41.1 50 12]{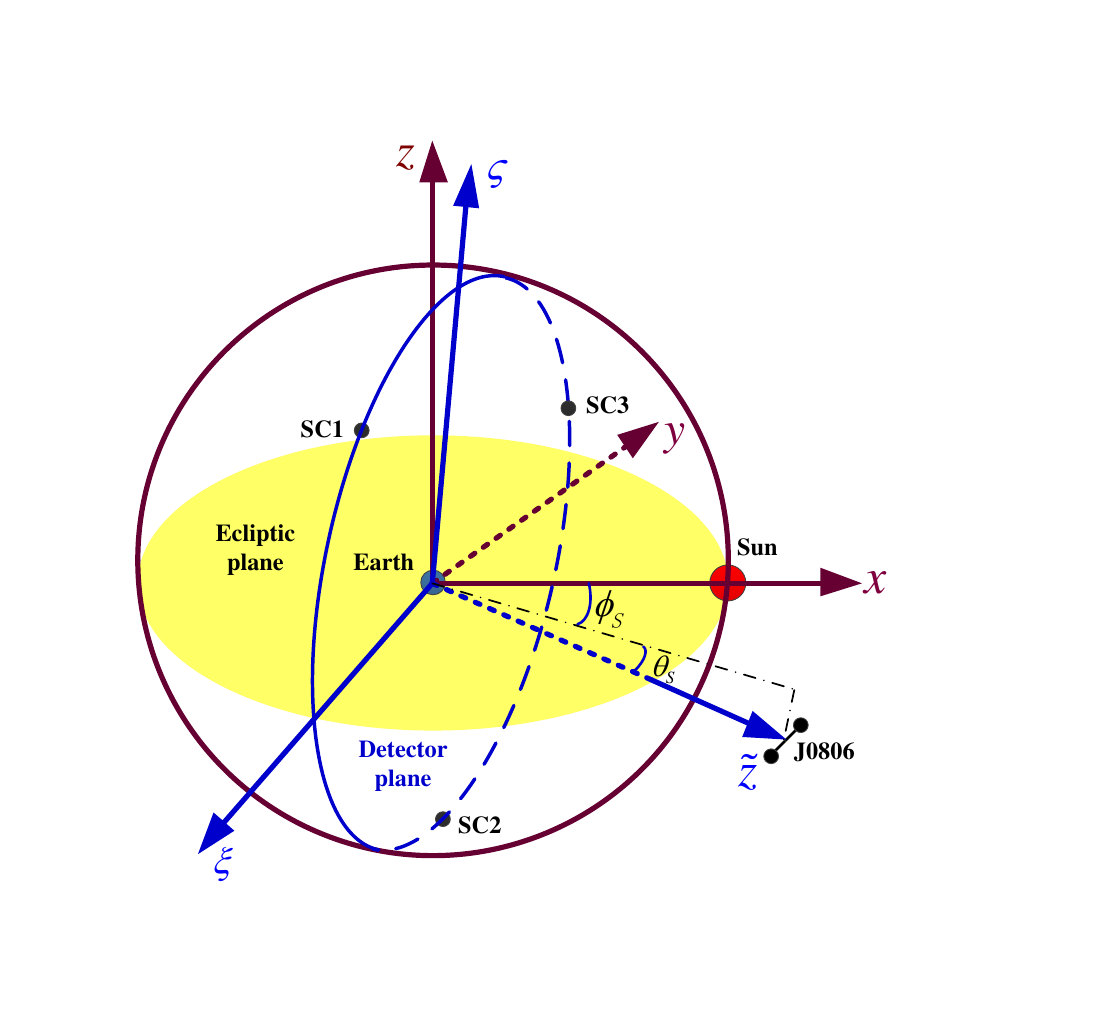}
	\caption{The schematic view of the detector plane in the GSE coordinates ($x$, $y$, $z$). $\tilde z$ is the normal of the detector's plane. $\theta_s$ is the angle between the projection of $\tilde z$ on the ecliptic plane and $\tilde z$. $\phi_s$ is the angle between the direction from the earth to the sun ($x$ axis) and the projection of $\tilde z$ on the ecliptic plane. $\xi$ is along the intersection of the detector's plane and the ecliptic plane, and $\zeta$ is along the intersection of a plane perpendicular to $\xi$ and the detector's plane. }
	\label{fig:02}
\end{figure}

\begin{figure}[!h]
	\centering
	\includegraphics[width=0.8\textwidth]{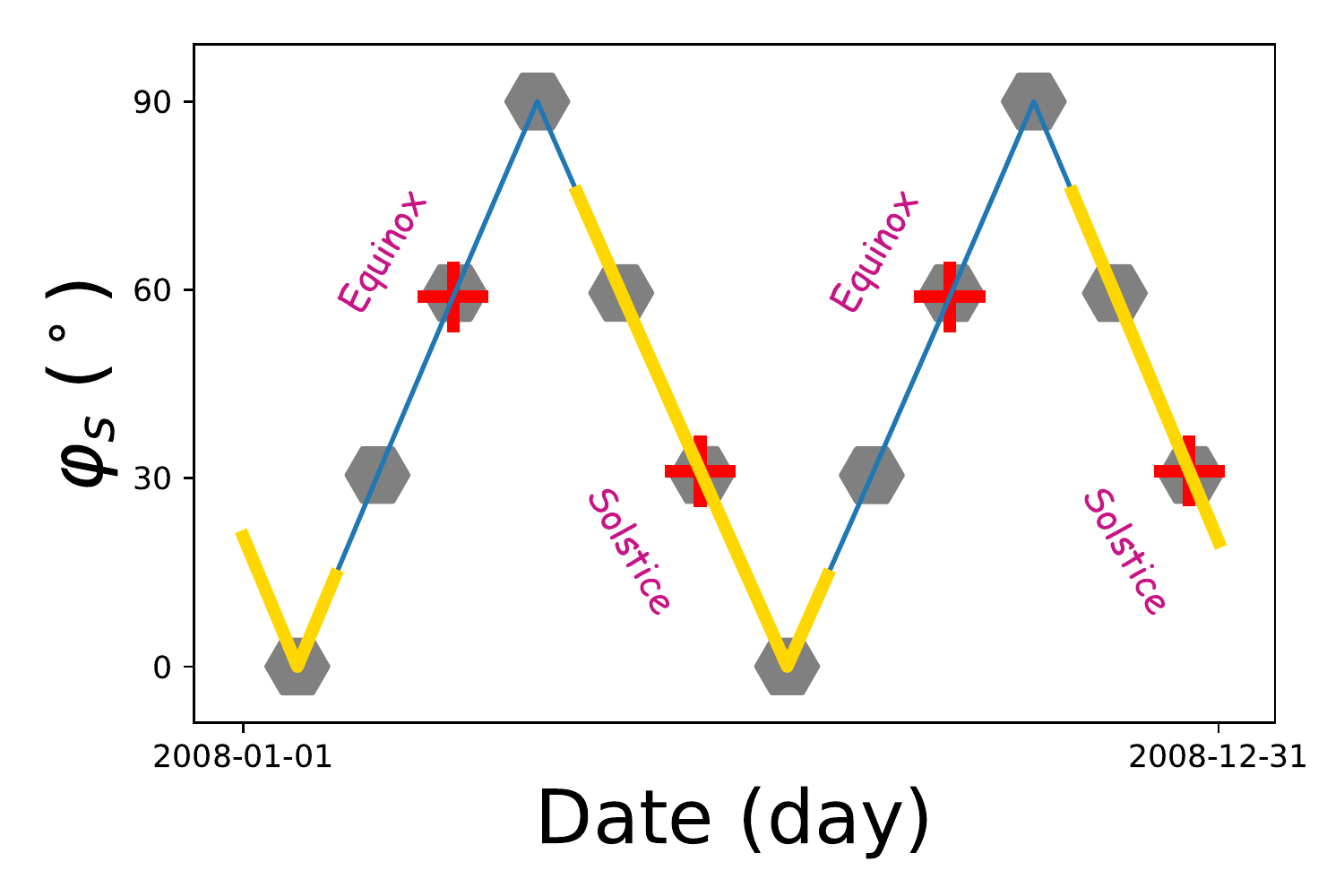}
	\caption{The time variation of $\varphi_{s}$ in the year 2008. The thick yellow lines represent the observation duration of TQ, the thin blue lines represent the non-observation duration of TQ. The red pluses mark the spring and the autumn equinoxes, the summer and the winter solstices. $\varphi_{s}$ on the grey hexagons are equal to 0$^\circ$, 30$^\circ$, 60$^\circ$ and 90$^\circ$, respectively. }
	\label{fig:1}
\end{figure}

The TQ detector's plane facing the reference source is approximately perpendicular to the ecliptic plane. 
The global geometric structure of the bow shock and the magnetopause are quasi-axisymmetric 
along the Sun-Earth line ($-x$ axis in the GSE coordinates). 
As shown in Fig. \ref{fig:02}, the normal of the detector's plane is denoted as $\tilde z$, 
$\xi$ is along the intersection of the detector's plane and the ecliptic plane, 
and $\zeta$ is along the intersection of a plane perpendicular to $\xi$ and the detector's plane. 
Since the angle between $\tilde z$ and the ecliptic plane is only 4.7$^{\circ}$, $\zeta$ is approximately 
perpendicular to the ecliptic plane.
The angle $\phi_{s}$ between the direction from the sun to the earth and the projection of 
$\tilde z$ on the ecliptic plane ranges from 0 to 360$^\circ$ annually, 
with $\phi_{s} = 120.5^\circ$ at the spring equinox \cite{Hu2018}. 
During one revolution of the TQ spacecraft around the Earth (3.65 days), 
$\phi_{s}$ can be approximately regarded  as a constant. 
In order to describe the relative position of the TQ detector's plane and the geometric structure of the earth magnetosphere conveniently, we transform $\phi_{s}$ to its associated acute angle $\varphi_{s}$ in the GSE coordinates. 
Taking the year 2008 as an example, numerical calculation based on JPL 
Planetary and Lunar Ephemeris DE421 \cite{DE421-2009}  is adopted to evaluate the time-varying $\varphi_{s}$ which is shown as the solid lines in Fig. \ref{fig:1}. The spring and autumn equinoxes, the summer and winter solstices are shown as red pluses. The observation windows of TQ is 2 $\times$ 3 months in one sidereal year \cite{Luo2016}. 
$\varphi_{s}$ in observation windows (thick yellow lines around summer and winter solstices) and non-observation windows (thin dark blue lines around spring and autumn equinoxes) range from 0$^{\circ}$ to 75.5$^{\circ}$ and  from 14.5$^{\circ}$ to 90$^{\circ}$, respectively. 

\begin{figure}[!h]
	\centering
	\includegraphics[width=0.9\textwidth]{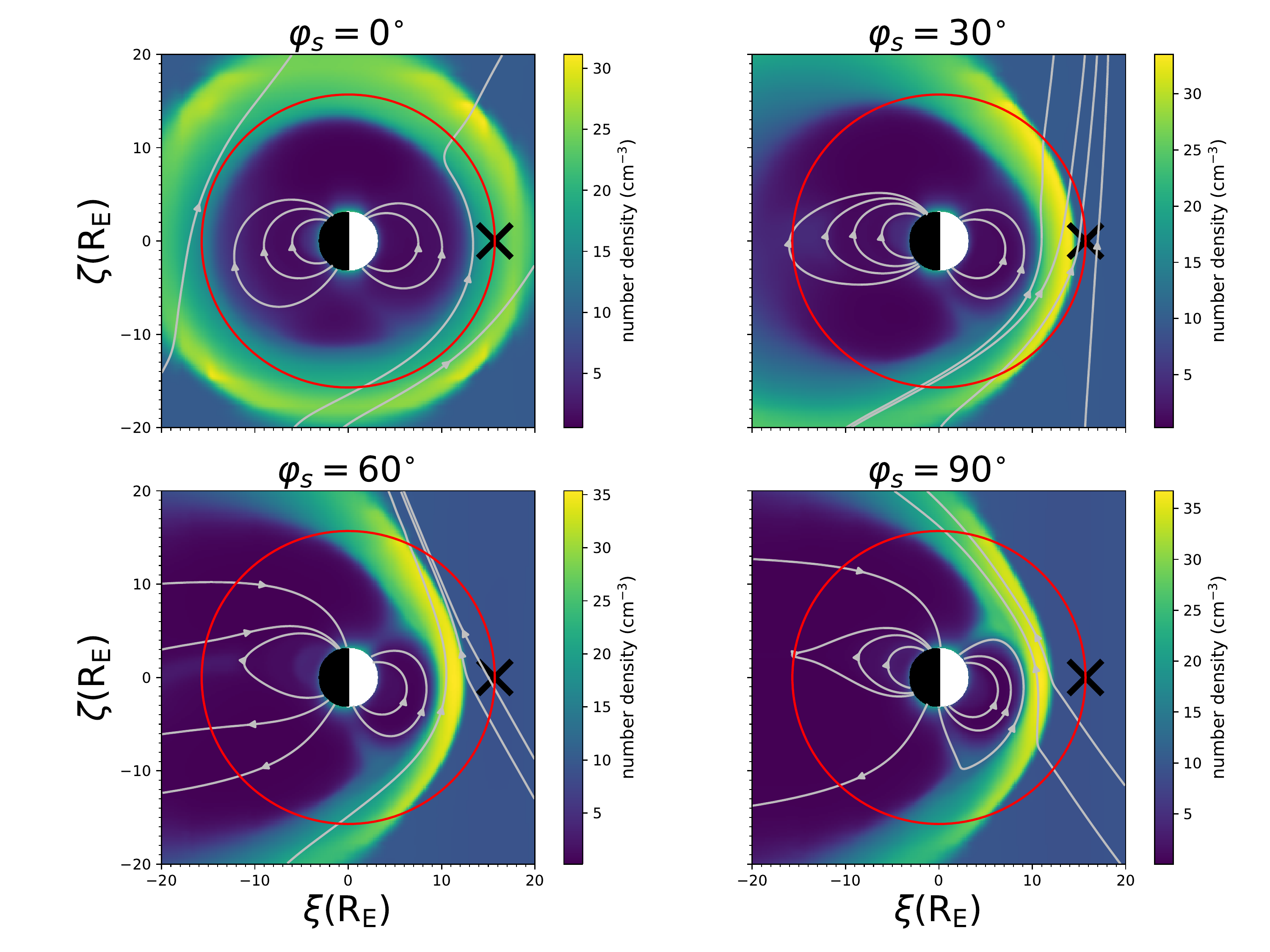}
	\caption{The electron number densities on the TQ's detector planes where $\varphi_{s}$ = 0$^\circ$, 30$^\circ$, 60$^\circ$ and 90$^\circ$, respectively. Black crosses mark the initial phase for one of the TQ's spacecraft. White lines are the representative magnetic field lines in the simulation domain. }
	\label{fig:2}
\end{figure}

In section~\ref{s:4}, we study the acceleration noises in four representative relative positions with $\varphi_{s}$ = 0$^{\circ}$, 30$^{\circ}$, 60$^{\circ}$, and 90$^{\circ}$ which are marked as grey hexagons in Fig. \ref{fig:1}. The hexagons on the yellow lines are in the observation window of TQ.  
The electron number densities for these four cases are given in Fig.~\ref{fig:2}, 
in which the simulation result for 2008-05-01 20:00 UT is used. 
The red circle represents the orbit of the spacecraft. 
The black crosses located at ($\xi = 15.7$, $\zeta = 0$) ${\rm R_{E}}$ mark the initial phase for one of the spacecraft. 
Two distinct boundaries around 10--20 ${\rm R_{E}}$ can be seen: the outer one is the bow shock, the inner one is the magnetopause. 
For $\varphi_{s}=0^{\circ}$, the bow shock and magnetopause are approximately axisymmetric, the bow shock is slightly larger than 20 ${\rm R_{E}}$, while the magnetopause is slightly larger than 10 ${\rm R_{E}}$. 
For $\varphi_{s} =  30^{\circ}$, $60^{\circ}$ and 90$^{\circ}$, the spacecraft will pass through the solar wind, the bow shock, the magnetosheath, the magnetopause, the lobes of magnetosphere and the magnetotail. We can see that as $\varphi_{s}$ decreasing the orbit will gradually shrink into the magnetosheath.

\section{Results}
\label{s:4}

\subsection{Space plasma and magnetic field} 
\label{s:4.1}

\begin{figure}[ht]
	\centering
		\begin{minipage}[b]{\textwidth}
			\centering
			\begin{minipage}[b]{0.24\textwidth}
				\includegraphics[width = 4.05 cm,trim = 0 0 0 0,clip = true]{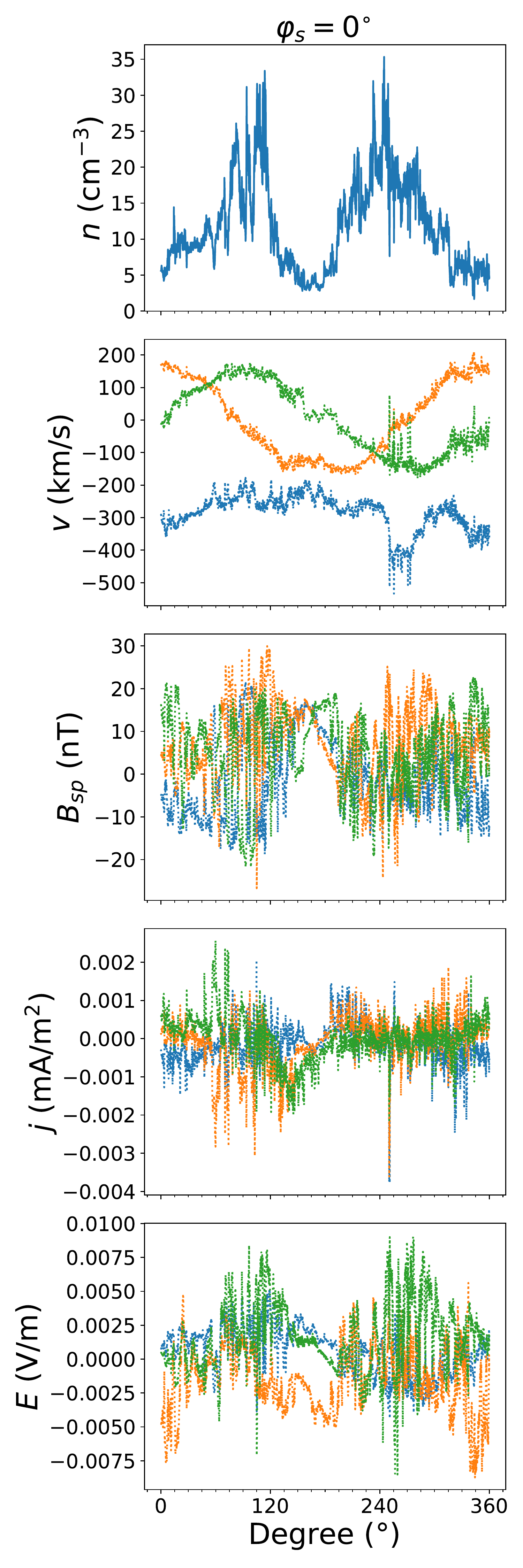}
			\end{minipage}
			\begin{minipage}[b]{0.24\textwidth}
				\includegraphics[width = 4.05 cm,trim = 0 0 0 0,clip = true]{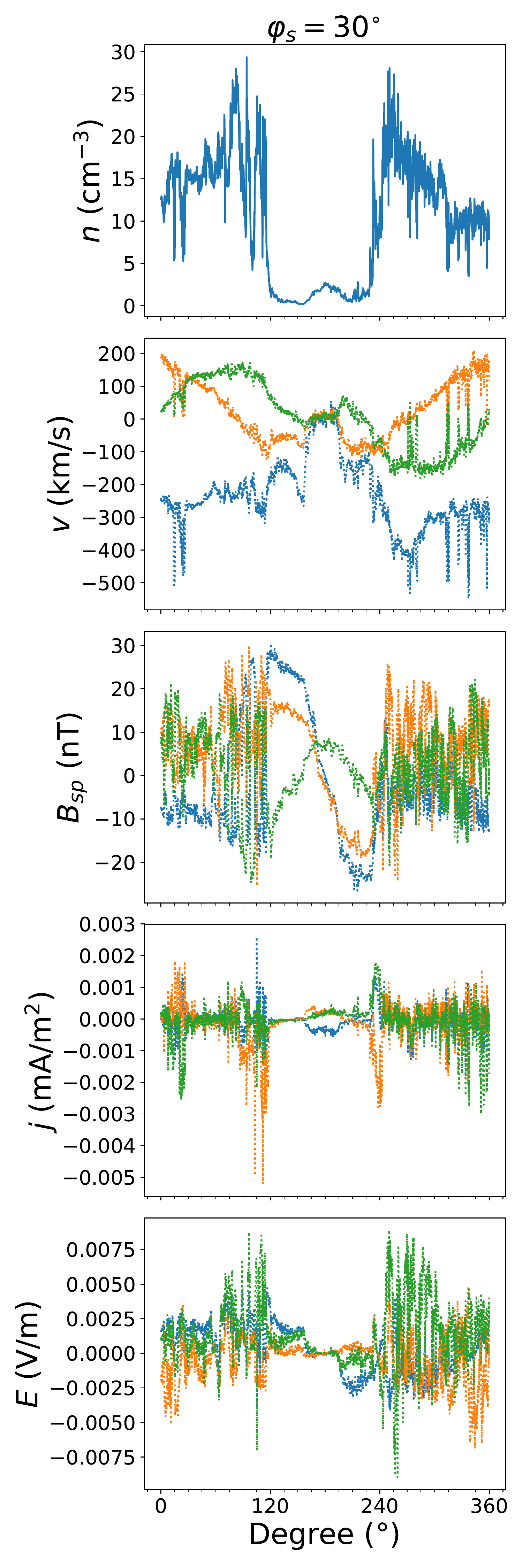}
			\end{minipage}
			\begin{minipage}[b]{0.24\textwidth}
				\includegraphics[width = 4.05 cm,trim = 0 0 0 0,clip = true]{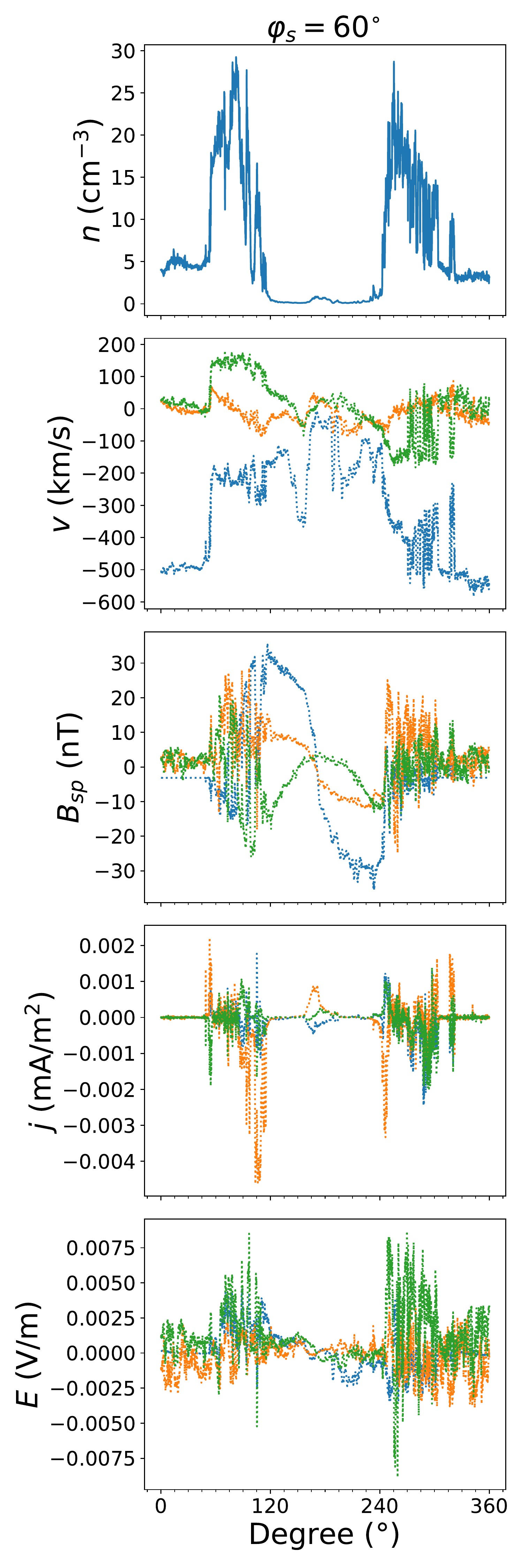}
			\end{minipage}
			\begin{minipage}[b]{0.24\textwidth}
				\includegraphics[width = 4.05 cm,trim = 0 0 0 0,clip = true]{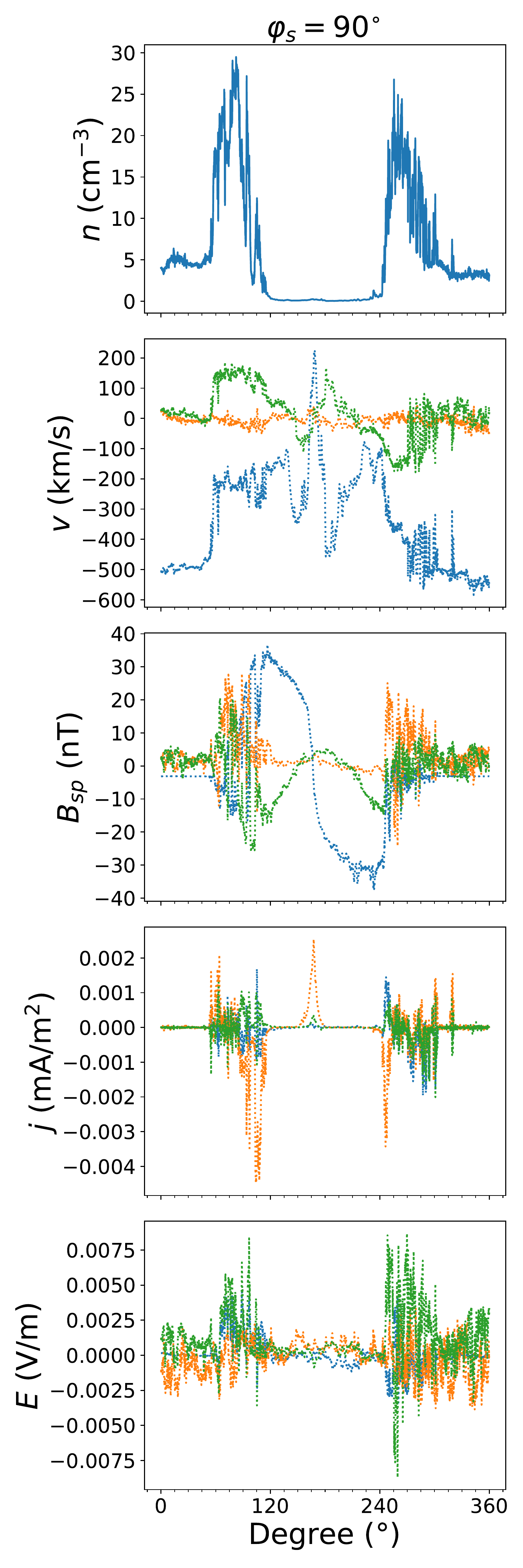}
			\end{minipage}
		\end{minipage}
	
	\caption{Distributions of ion number density $n_i$, bulk flow $\bm{v}$, space magnetic field $\bm{B}_{\rm sp}$, electric current $\bm{J}$ and electric field $\bm{E}_{\rm sp}$ on the orbit are in rows 1 to 5, respectively. The blue, orange and green lines in rows 2 to 5 represent the $x$, $y$, $z$ components of $\bm{v}$, $\bm{B}_{\rm sp}$, $\bm{J}$, and $\bm{E}_{\rm sp}$ in the GSE coordinates. Columns 1 to 4 denote the distributions of these parameters on the detector's plane where $\varphi_{s}$ = 0$^\circ$, 30$^\circ$, 60$^\circ$ and 90$^\circ$. Note that the abscissa represents the 
	orbital phase for one spacecraft in a 3.65-day circular orbit around the Earth.  }
	\label{fig:3}
\end{figure}

On the spacecraft orbit, the values of the space plasma and magnetic field parameters 
are obtained from the interpolation of the ones on the 3D grid of the simulation domain. 
Here, the position of each spacecraft is sampled every 60 s which is in line with the time resolution of the simulation. 
For $\varphi_{s} = 0^{\circ}, 30^{\circ}, 60^{\circ}$ and $90^{\circ}$, 
the number density of ion $n_i$, bulk flow $\bm{v}$, magnetic field $\bm{B}_{\rm sp}$, electric current $\bm{J}$ 
and electric field ${\bm E}_{\rm sp}$ on a spacecraft's orbit are shown in Fig.~\ref{fig:3}. 
The dotted blue, orange and green lines represent the $x$, $y$ and $z$ components of $\bm{v}$, $\bm{B}_{\rm sp}$, $\bm{J}$ and ${\bm E}_{\rm sp}$, respectively. 
Note that ${\bm E}_{\rm sp}$ is calculated from $- \bm{v} \times \bm{B}_{\rm sp}$ as mentioned in section \ref{s:2.3}.

Take $\varphi_{s} = 90^{\circ}$ as an example, combining Fig.~\ref{fig:2} and Fig.~\ref{fig:3}, 
we can see that $n_i$ in the solar wind 
is lower than that in the magnetosheath, but larger than that in the magnetosphere. Meanwhile, the absolute value of $v_x$ is obviously larger than that of $v_y$ and $v_z$, the absolute value of $\bm{B}_{\rm sp}$ is smaller than those in the magnetosheath and magnetosphere. 
In the magnetosheath, the absolute values of $n_i$ and $\bm{B}_{\rm sp}$ are larger than those in solar wind, because that the magnetosheath is the downstream of the bow shock, $n_i$ and $\bm{B}_{\rm sp}$ can be enhanced by the shock \cite{Draine1993,Su2016}; and the fluctuations of all these parameters are larger than those in the solar wind and the  magnetosphere.
In the magnetosphere, $n_i$ is lower than these in the solar wind and the magnetosheath, the absolute values of $\bm{B}_{\rm sp}$ is larger than that in the solar wind. 
In the magnetotail,  the $x$ component of $\bm{B}_{\rm sp}$ reverses and the absolute value of $\bm{J}$ becomes larger, since there is a cross-tail current sheet in the magnetotail \cite{Runov2011}.
All these features are consistent with the observations \cite{Dimmock2013,Wang2018}. 
As $\varphi_{s}$ decreases, the TQ spacecraft will spend more time in the magetosheath 
and the time series of these parameters become more fluctuated.

\subsection{Residual accelerations}
\label{s:4.2}

\begin{figure}[ht]
	\centering
	\begin{minipage}[b]{\textwidth}
		\centering
		\begin{minipage}[b]{0.24\textwidth}
			\includegraphics[width = 4.05 cm,trim = 0 0 0 0,clip = true]{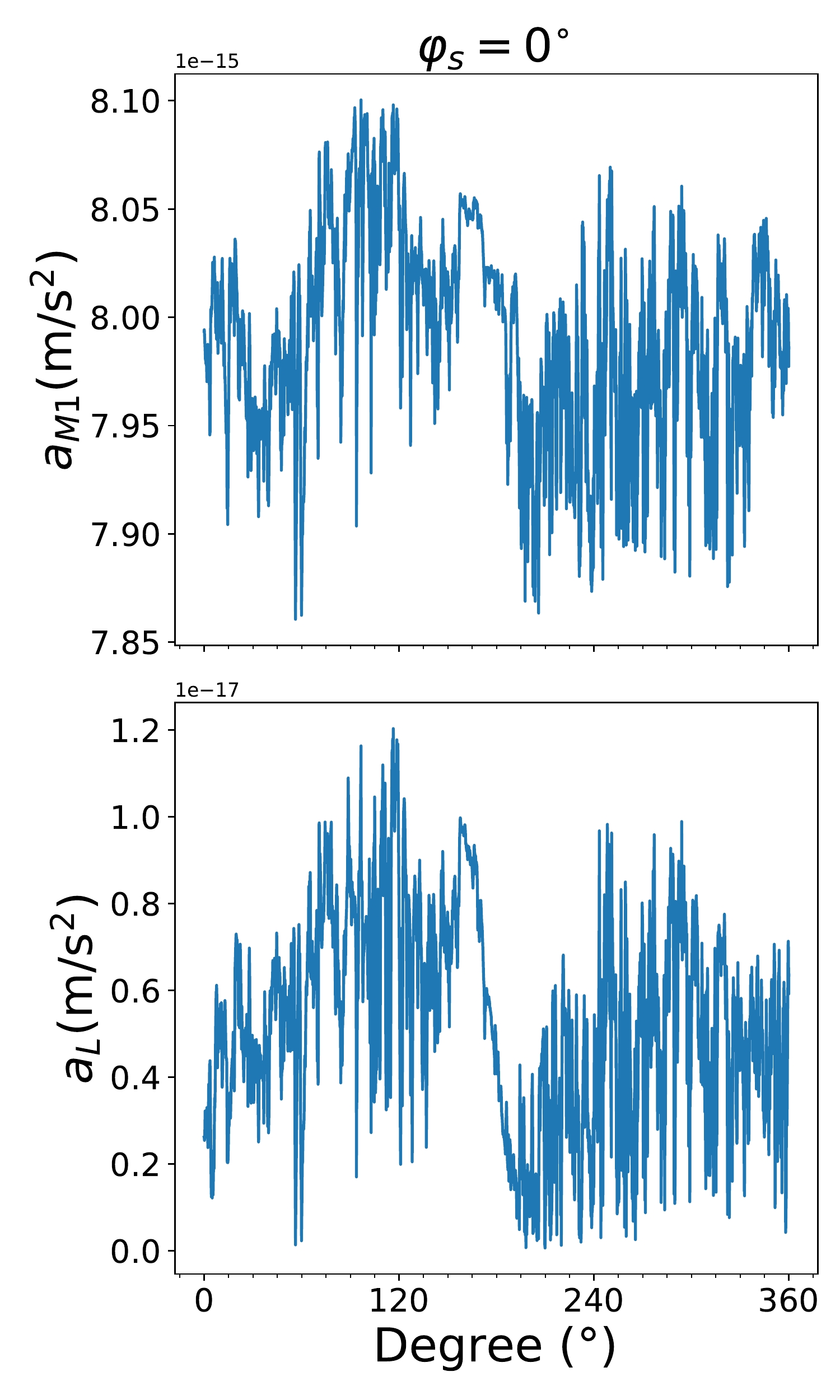}
		\end{minipage}
		\begin{minipage}[b]{0.24\textwidth}
			\includegraphics[width = 4.05 cm,trim = 0 0 0 0,clip = true]{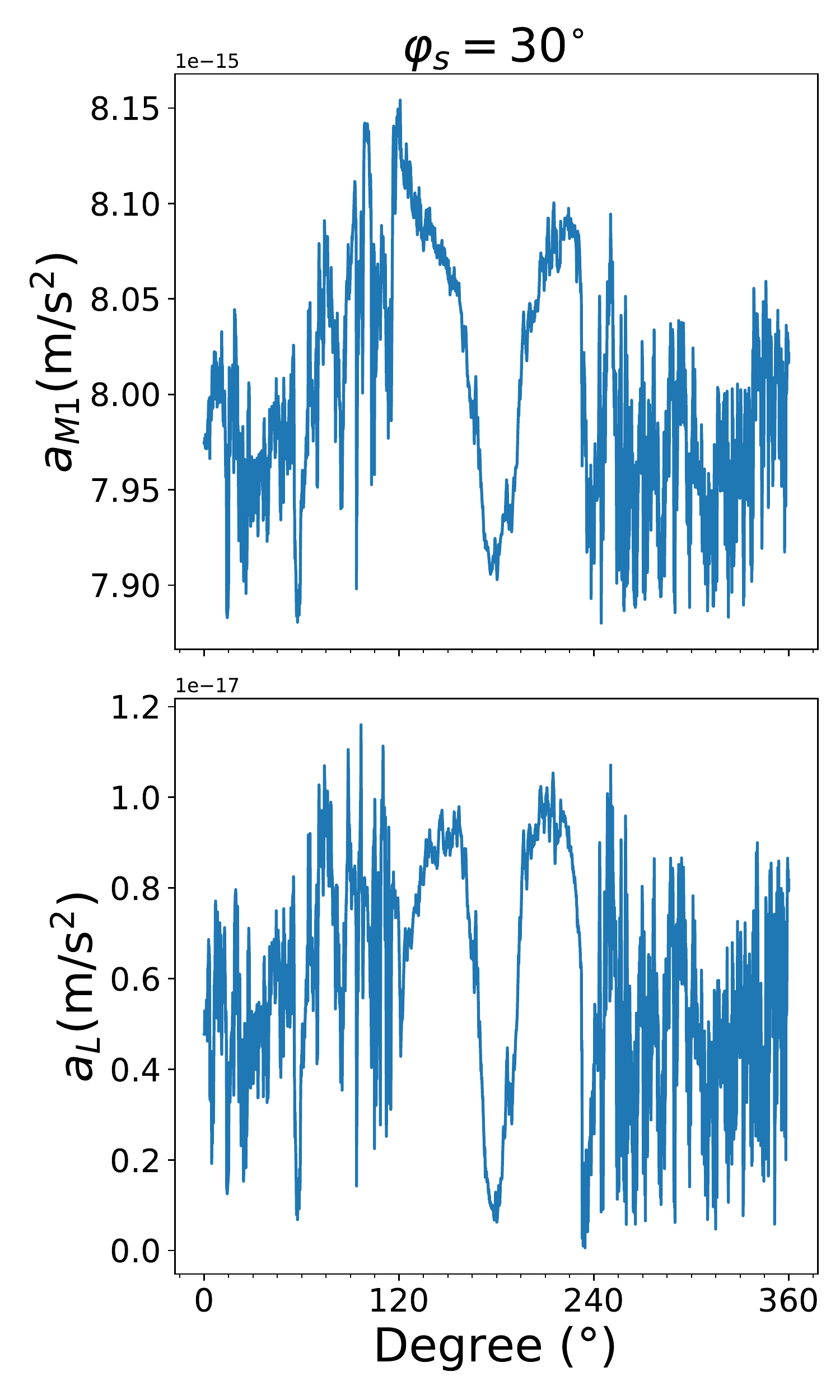}
		\end{minipage}
		\begin{minipage}[b]{0.24\textwidth}
			\includegraphics[width = 4.05 cm,trim = 0 0 0 0,clip = true]{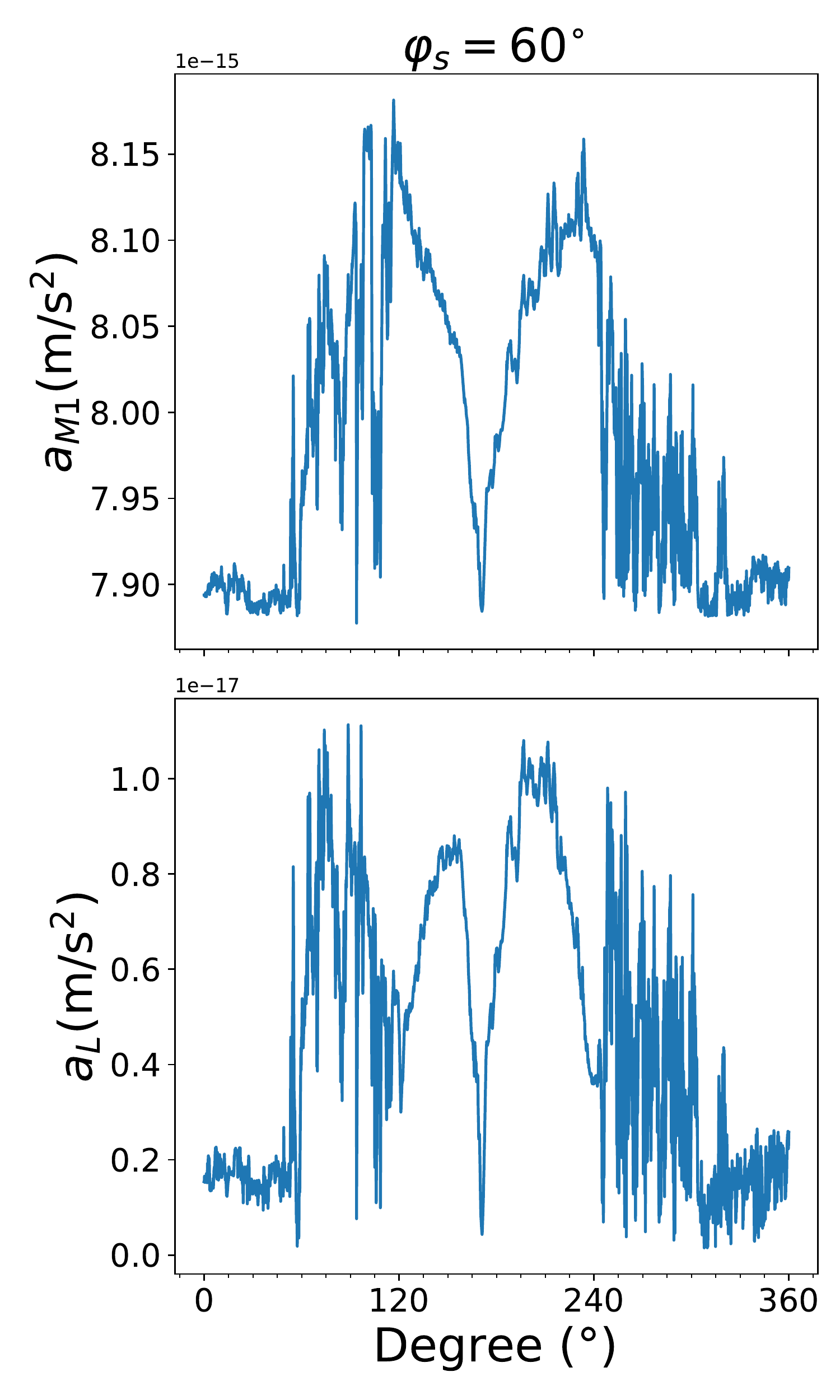}
		\end{minipage}
		\begin{minipage}[b]{0.24\textwidth}
			\includegraphics[width = 4.05 cm,trim = 0 0 0 0,clip = true]{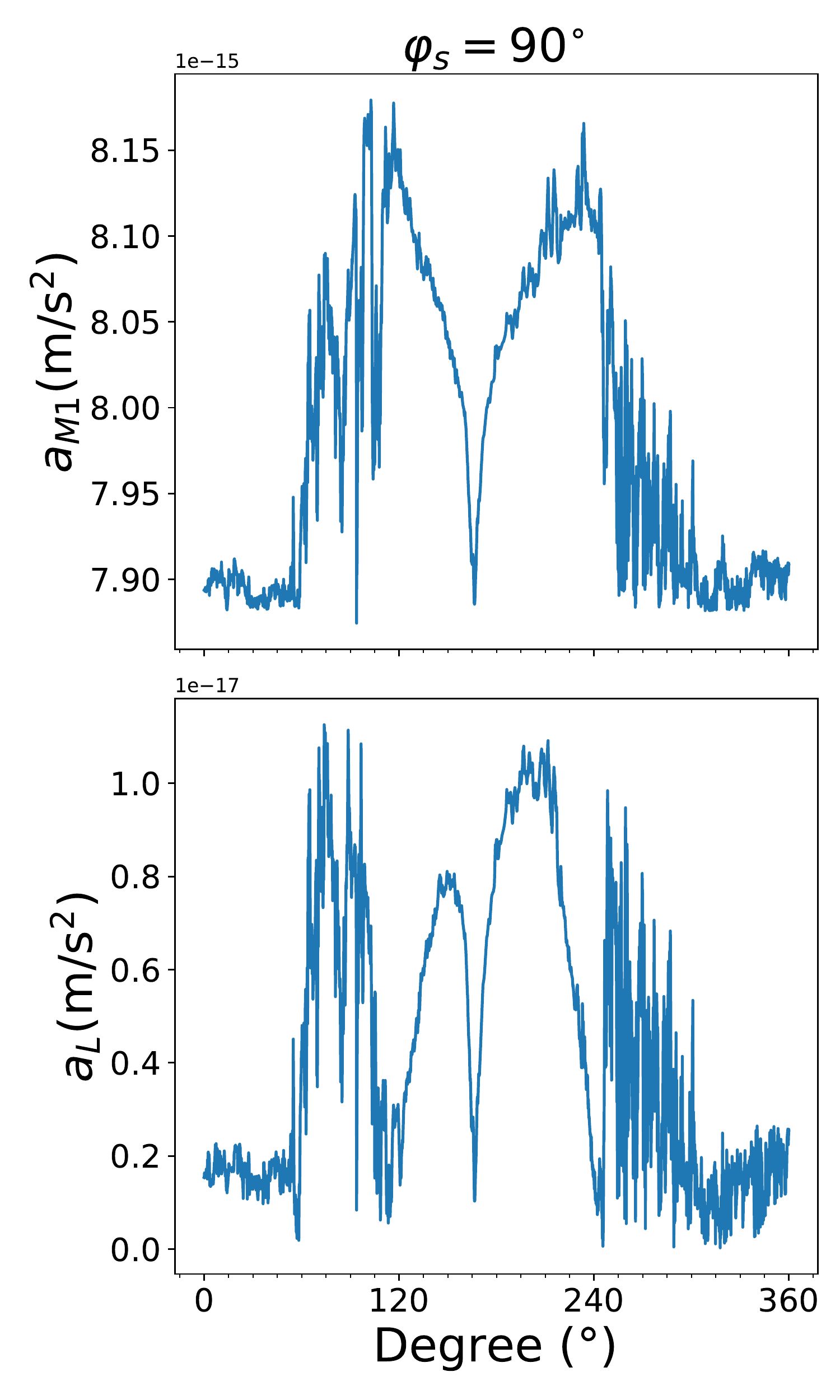}
		\end{minipage}
	\end{minipage}
	
	\caption{Distributions of $\bm{a}_{\rm M1}$ and $\bm{a}_{\rm L}$ 
	at $\varphi_{s}$ = 0$^\circ$, 30$^\circ$, 60$^\circ$ and 90$^\circ$. }
	\label{fig:4}
\end{figure}

According to Equation (\ref{equation11})--(\ref{equation13}) and 
the magnitudes of $\bm{B}_{\rm sp}$ and $\bm{E}_{\rm sp}$ on the orbit, 
the amplitudes of $\bm{a}_{\rm M1}$, $\bm{a}_{\rm M2}$, $\bm{a}_{\rm M3}$, $\bm{a}_{\rm M4}$ 
and $\bm{a}_{\rm L}$ are of the order of $10^{-14}$, $10^{-20}$, $10^{-29}$, $10^{-22}$ 
and $10^{-17}$ $\mathrm{m/s^2}$, respectively. 
$\bm{a}_{\rm M1}$ is the dominant acceleration noise caused by the space plasma. 
The time series of the acceleration noises of $\bm{a}_{\rm M1}$ and $\bm{a}_{\rm L}$ are shown in Fig. \ref{fig:4}.
Since $\bm{M}_{\rm r} \approx 20~ \mathrm{nA~m^2}$, $\bm{M}_{\rm isc} \approx 16~\mathrm{nA~m^2}$, $\bm{M}_{\rm isp} \approx 0.2~ \mathrm{nA~m^2}$, 
$\bm{M}_{\rm r} + 2 \bm{M}_{\rm isp}$ and $\bm{M}_{\rm r} + 2 \bm{M}_{\rm isc}$
are of the same order of magnitude, 
the six orders of magnitude difference between $\bm{a}_{\rm M1}$ and $\bm{a}_{\rm M2}$ 
is mainly due to the difference between the values of 
$\nabla \bm{B}_{\rm sc}$ and $\nabla \bm{B}_{\rm sp}$.

In addition, the time fluctuations of $\bm{B}_{\rm sc}$, $\nabla \bm{B}_{\rm sc}$ and 
$\bm{E}_{\rm sc}$ (denoted as $\delta \bm{B}_{\rm sc}$, $\delta \nabla \bm{B}_{\rm sc}$ 
and $\delta \bm{E}_{\rm sc}$)  will result in the residual accelerations in $\bm{a}_{\rm M5}$ ($\delta \bm{E}_{\rm sc}$), 
$\bm{a}_{\rm M6}$ ($\delta \nabla \bm{B}_{\rm sc}$) and an additional term  
in $\bm{a}_{\rm M1}$ ($\delta \nabla \bm{B}_{\rm sc}$, hereafter denoted as $\bm{a}'_{\rm M1}$). 
The typical value of $\delta\nabla \bm{B}_{\rm sc}$ is 10 $\rm nT/m/Hz^{1/2}$ at 0.1 mHz 
with the frequency dependence of $1/f^{3/2}$ \cite{Stebbins2004}; 
and $\delta\bm{E}_{\rm sc}$ is 100 $\rm \mu V/Hz^{1/2}$ at 1 mHz \cite{LPF2011}. 
The magnitude of $\bm{a}'_{\rm M1}$, $\bm{a}_{\rm M5}$ and $\bm{a}_{\rm M6}$ can be therefore 
estimated as $10^{-19}$, $10^{-32}$, and $10^{-18}$ $\rm m/s^2/Hz^{1/2}$ at 1 mHz, respectively, 
which will be ignored in the following analysis.

For the MHD simulation, the ratio between displacement current density and electric current density can be approximated as follow \cite{ChenFF2016,Swanson2003}: 
\begin{equation}
\label{equation17}
\frac{\varepsilon_0 \partial \bm{E}_{\rm sp}/\partial t}{\bm{j}} \sim \frac{E/T}{c^2 B/L} \sim \frac{v \times B L/T}{c^2 B} \sim \frac{v^2}{c^2},
\end{equation}
where $T$, $L$ and $v$ are the typical time, length, and velocity of MHD bulk flow in the magnetosphere and the solar wind, respectively. Here $v \sim 10^2$ km/s in the magnetosphere or the solar wind at 1 AU. 
From Equation (\ref{equation17}), we can see that the displacement current density is much lower than the electric current density. 
On the contrary, for the electromagnetic (EM) waves in plasma,  which are ubiquitous in heliophysics \cite{Huang2016,Zhao2015,Zhao2017}, 
the displacement current density and the electric current density are approximately on the same order of magnitude. 
However, these EM waves can not be revealed in the MHD simulation. 
The impact of EM waves in plasma, especially on $\bm{a}_{\rm M3}$, will be subjected to our future investigations.

\begin{figure}[ht]
	\centering
	\begin{minipage}[b]{\textwidth}
		\centering
		\begin{minipage}[b]{0.24\textwidth}
			\includegraphics[width = 4.05 cm,trim = 0 0 0 0,clip = true]{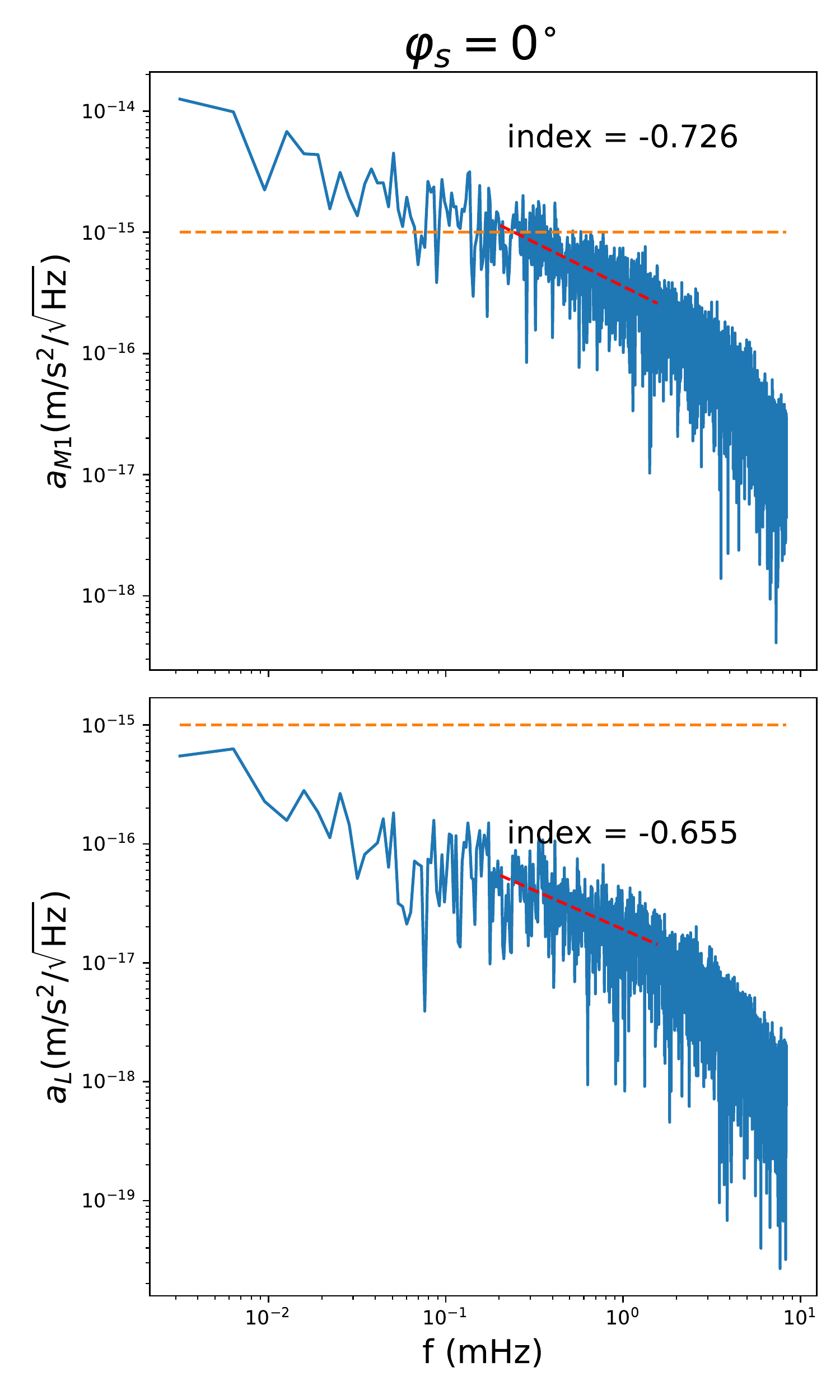}
		\end{minipage}
		\begin{minipage}[b]{0.24\textwidth}
			\includegraphics[width = 4.05 cm,trim = 0 0 0 0,clip = true]{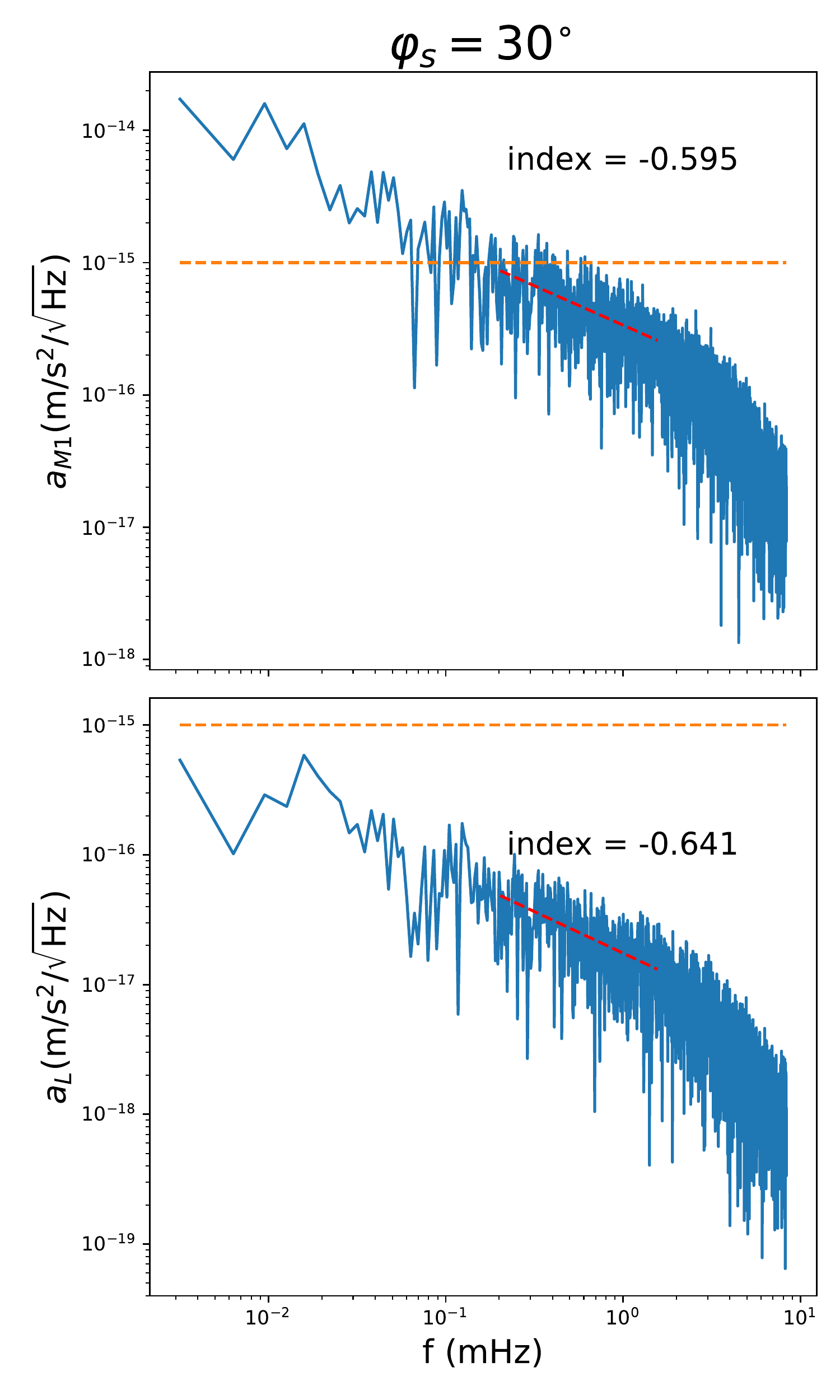}
		\end{minipage}
		\begin{minipage}[b]{0.24\textwidth}
			\includegraphics[width = 4.05 cm,trim = 0 0 0 0,clip = true]{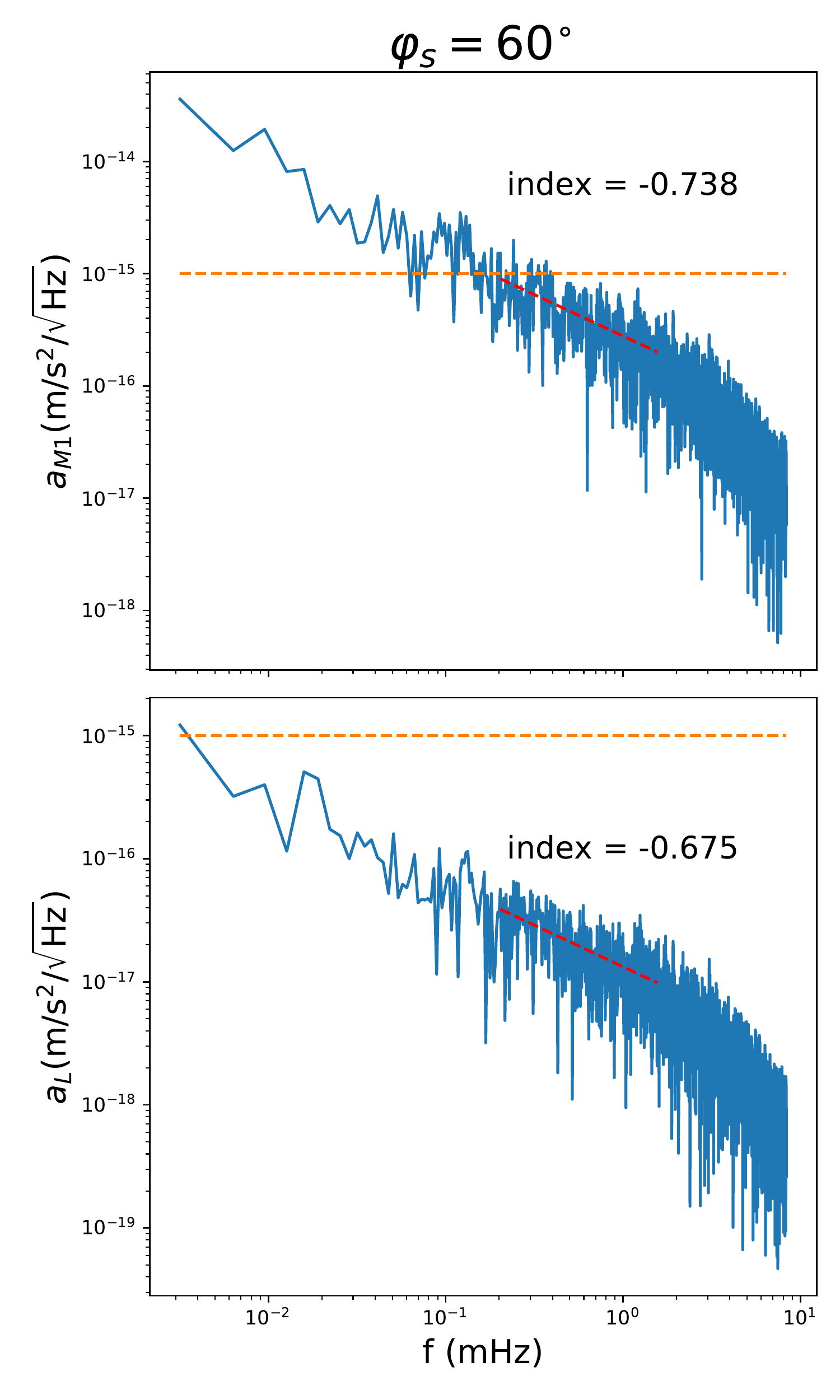}
		\end{minipage}
		\begin{minipage}[b]{0.24\textwidth}
			\includegraphics[width = 4.05 cm,trim = 0 0 0 0,clip = true]{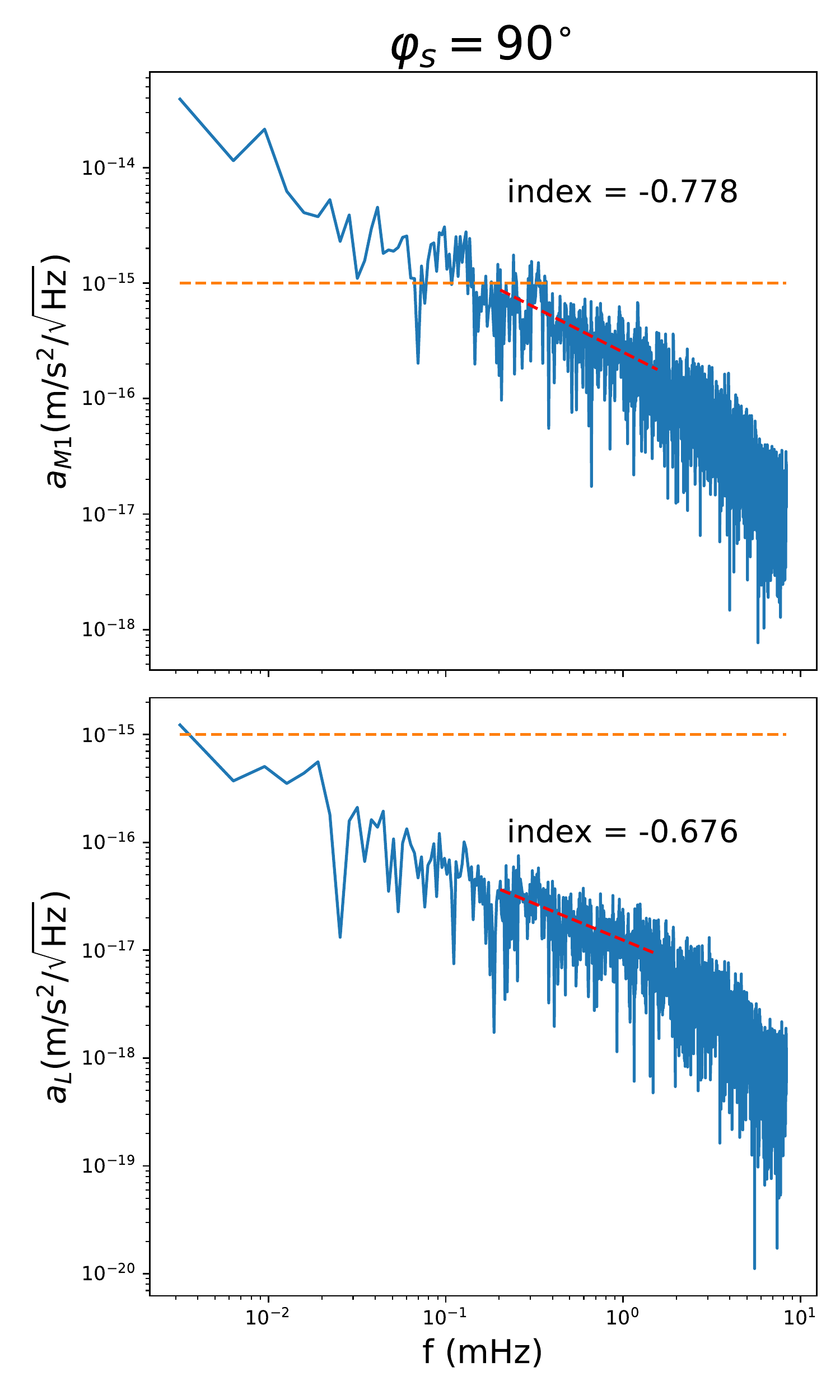}
		\end{minipage}
	\end{minipage}
	
	\caption{Amplitude spectral densities of $\bm{a}_{\rm M1}$ and $\bm{a}_{\rm L}$ at $\varphi_{s}$ = 0$^\circ$, 30$^\circ$, 60$^\circ$ and 90$^\circ$. 
The orange dashed horizontal lines mark the preliminary goal of the acceleration noise for TQ. 	
The red dashed line mark the best fits of the ASDs of $\bm{a}_{\rm M1}$ and $\bm{a}_{\rm L}$. }
	\label{fig:5}
\end{figure}

Fig. \ref{fig:5} shows the amplitude spectral densities (ASDs) of the time series for $\bm{a}_{\rm M1}$ and $\bm{a}_{\rm L}$. 
The dashed horizontal lines ($\sqrt{S_{a}}=10^{-15}$ m/s$^2$/Hz$^{1/2}$) mark the 
preliminary goal of the acceleration noise from the low-noise inertial sensor of TQ \cite{Luo2016}. 
The largest contribution to the ASD of the acceleration noise is that of $\bm{a}_{\rm M1}$, 
while the ASD of $\bm{a}_{\rm L}$ approaches
$10^{-15}$ m/s$^2$/Hz$^{1/2}$ 
at the lowest frequency (1/3.65 day).

The Savitzky-Golay filter \cite{SavGol1964} is used here to smooth the ASDs 
of $\bm{a}_{\rm M1}$ and $\bm{a}_{\rm L}$ before 
the single power law fit. 
The fitting results are shown as the red dashed lines in Fig. \ref{fig:5}. 
We can see that the lower frequency parts of the best fit spectra for $\bm{a}_{\rm M1}$ approach 
$10^{-15}$ m/s$^2$/Hz$^{1/2}$ when $f \approx$ 0.24, 0.16, 0.18 and 0.17 mHz 
for $\varphi_{s}$ = 0$^{\circ}$, 30$^{\circ}$, 60$^{\circ}$ and 90$^{\circ}$, respectively. 
The corresponding spectral indexes 
for $\bm{a}_{\rm M1}$ ($\bm{a}_{\rm L}$) are -0.726 (-0.655), -0.595 (-0.641), -0.738 (-0.675) and -0.778 (-0.676).

Since the ASD of the acceleration noise caused by the space magnetic field ($\sqrt{S_{as}}$) is dominated 
by $\bm{a}_{\rm M1}$, we set $\sqrt{S_{as}} = 10^{-15}$ m/s$^2$/Hz$^{1/2}$ at 0.24 mHz and simply approximate 
the spectral index as $-2/3$. The ratios $\sqrt{S_{as}/S_a}$ are 0.38 and 0.08 at 1 mHz and 10 mHz, respectively, 
for the nominal values of the magnetic susceptibility ($\chi_{\rm m} = 10^{-5}$) and 
the magnetic shielding factor ($\xi_{\rm m} = 10$) of the test mass.

\begin{figure}[!h]
	\centering
	\includegraphics[width=0.75\textwidth]{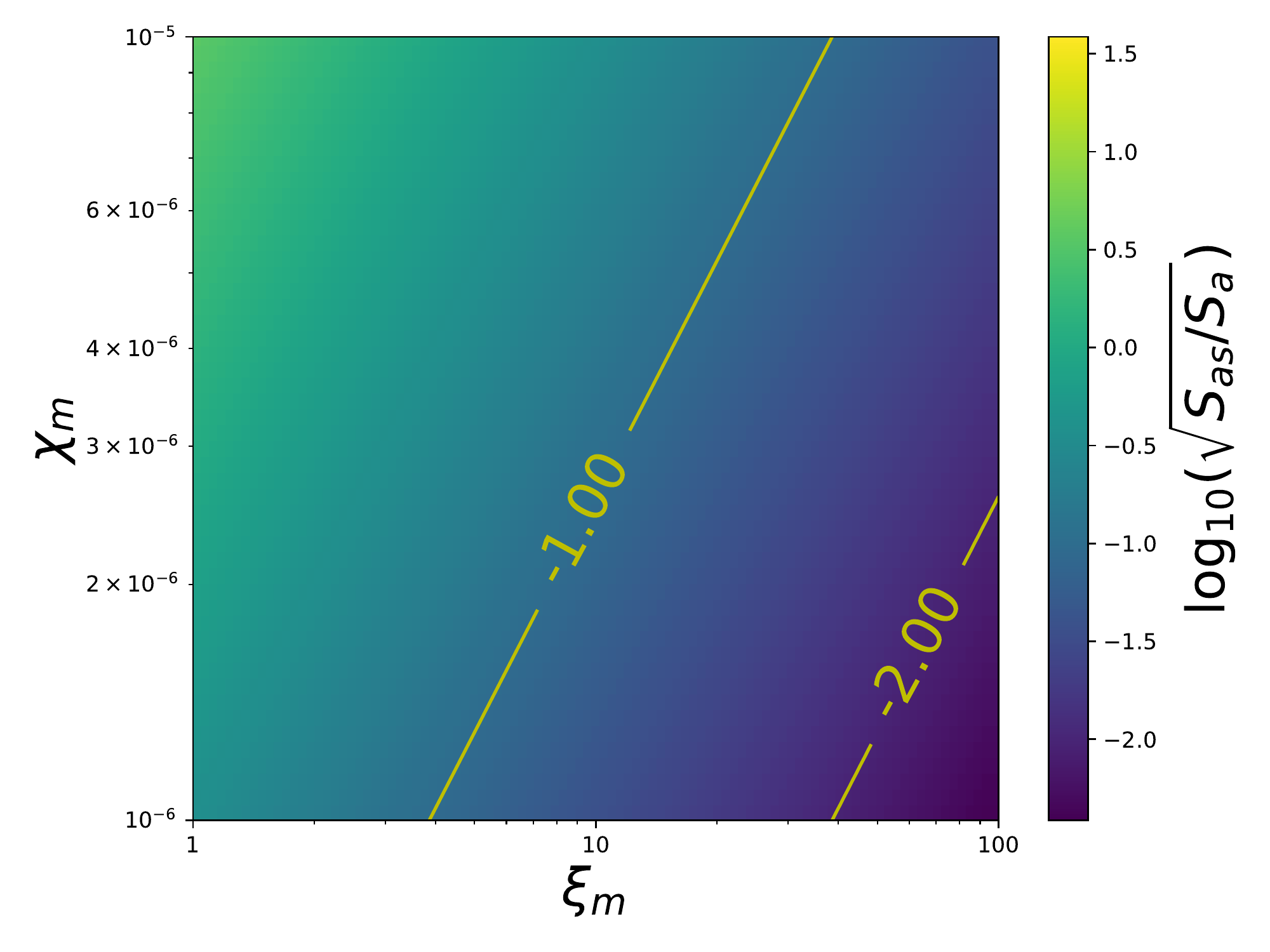}
	\caption{ $\sqrt{S_{as}/S_a}$ at 1 mHz in the parameter space of $\xi_{\rm m}-\chi_{\rm m}$. 
	The contours of $\sqrt{S_{as}/S_a} = 10^{-1}$ and $10^{-2}$ are shown as yellow lines. 
		 }
	\label{fig:6}
\end{figure}

Note that $\xi_{\rm m}$ and $\chi_{\rm m}$ are tuneable. 
For LISA, the proposed value of $\xi_{\rm m}$ is of the order of
10-20 \cite{Schumaker2002}, 
a factor of 10 larger than that of LISA Pathfinder \cite{Schumaker2003}. It can be further increased to 40 in 
the double magnetic shielding scheme \cite{Cavalleri_2009}; 
the value of $\chi_{\rm m}$ is proposed to be $10^{-6}$ \cite{Schumaker2003},
$3 \times 10^{-6}$ \cite{Stebbins2004} and $4 \times 10^{-6}$ \cite{Hanson2003}. 
Therefore, the aforementioned residual accelerations from space plasma can be regarded as an upper bound.
It will be further reduced by increasing $\xi_{\rm m}$ and/or decreasing $\chi_{\rm m}$. For example, shielding TMs with multiple shields and/or with novel materials can possibly increase $\xi_{\rm m}$ several times \cite{Cavalleri_2009}. On the other hand, the ultra-low $\chi_{\rm m}$ material for fabricating the TMs can be achieved by alloying diamagnetic and paramagnetic materials with a certain proportion \cite{Schumaker2003}. Besides, $\chi_{\rm m}$ is expected to be further minimized by optimizing the fabrication process such as increasing the alloying homogeneity and avoiding the introduction of impurities. 
As a reference, Fig. \ref{fig:6} gives $\sqrt{S_{as}/S_{a}}$ at 1 mHz in the parameter space of $\xi_m-\chi_m$. 
The yellow lines mark the contours of $\sqrt{S_{as}/S_{a}} = 10^{-1}$ and $10^{-2}$. 
We can see that $\sqrt{S_{as}/S_{a}}$ can be readily suppressed below  $10^{-1}$ 
by a mild adjust from the nominal values.

\section{Discussion and conclusions}
\label{s:5}

In this work, we study the acceleration noises caused by the magnetic field of space plasma for 
the test mass of TQ, 
which include the contributions from the magnetic force and Lorentz force. 
The SWMF is adopted to simulate the global structure of the earth magnetosphere. 
The solar wind conditions from the ACE observations with time resolution of 60 s are taken as inputs.  
The MHD simulation outputs the space plasma and magnetic field parameters which are  
then used to calculate the acceleration noises of the test mass on the orbit of TQ's spacecraft at 
$\varphi_{s} = 0^{\circ}, 30^{\circ}, 60^{\circ}$ and $90^{\circ}$. 
It turns out that the acceleration noise produced by the interaction between 
the TM's magnetic moment induced by the space magnetic field 
and the spacecraft magnetic field ($\bm{a}_{\rm M1}$) 
is the largest component with 
the ASD $\sqrt{S_{as}} \approx 10^{-15}$ m/s$^2$/Hz$^{1/2}$ at $f \approx 0.2$ mHz  
for the nominal values of $\xi_{\rm m}$ and $\chi_{\rm m}$. 
$\sqrt{S_{as}/S_a}$ are 0.38 and 0.08 at 1 mHz and 10 mHz, respectively. 
We further discuss the reduction of $\sqrt{S_{as}}$ in the parameter space of 
$\xi_{\rm m}-\chi_{\rm m}$ which can be considered as a reference of future instrumentation development for TQ. 
Note that the results obtained in Sec. \ref{s:4.2} depend on the values of 
the magnetic shielding factor and the magnetic susceptibility of the TM, 
for which we set  $\xi_{\rm m} = 10$ and $\chi_{\rm m} = 10^{-5}$, respectively, 
as fiducial values in this work. 
However, in reality, $\xi_{\rm m}$ and $\chi_{\rm m}$ will be subjected to 
the limits from engineering consideration (build, test and launch of the spacecraft) and metallurgy. 
Thus, it is crucial to verify the feasibility and robustness of $\xi_{\rm m}$ and $\chi_{\rm m}$  
in the future development of TQ.

The temporal resolution of our simulation is 60 s (corresponding to a Nyquist frequency of 1/120 Hz), therefore 
the phenomena with dynamic frequencies higher than about $10^{-2}$ Hz will not be visible. 
It is certainly important and will be subjected to our future investigations to study the space magnetic field 
with the aid from  high-temporal high-spacial resolution observations and simulations. 

On the other hand, the results presented here are based on the simulation from 
the SWMF which is a MHD model. 
Thus, neither the phenomena in the plasma scale, such as plasma waves and turbulences \cite{Howes2012,Salem2012,Huang2010,Huang2016}, nor the associated residual accelerations can 
be revealed in the current work. 
Furthermore, in the solar wind inputs, $P_{\rm dyn}$ 
owns moderate values with the mean of $\approx 2.1$ (see the bottom panel of Fig.~\ref{fig:01}).  
However, this value can be increased significantly in the eruption events, such as ICMEs, 
IP shocks, magnetic reconnections, etc. 
The impacts of these on the TM's residual acceleration and the spacecraft per se will be followed up.

\section*{Acknowledgements}
Simulation results have been provided by the Community Coordinated Modeling Center (CCMC) at Goddard Space Flight Center through their public Runs on Request system. This work was carried out using the SWMF and BATSRUS tools developed at the University of Michigan's Center for Space Environment Modeling. The modeling tools are available online through the University of Michigan for download and are available for use at CCMC. We are grateful to J.S. Zhao for valuable discussion on the plasma theory. This work is supported by NSFC under grants 11803008, Jiangsu NSF under grants BK20171108. W.Y. is supported by NSFC under grants 11973024, 11690021, 91636111 and `the Fundamental Research Funds for the Central Universities' under Grant No. 2019kfyRCPY106. T. Shi is supported by NASA grant 80NSSC17K0453. G.Y. is supported by NSFC under grants 11773016, 11533005, and 11961131002.
We thank the anonymous referees for helpful comments and suggestions. 

\section*{References}

\vspace{2 cm}

\end{document}